\PassOptionsToPackage{hyphens}{url}
\PassOptionsToPackage{svgnames}{xcolor}

\documentclass[compsoc,conference,a4paper,10pt,times]{IEEEtran}

\usepackage{amsmath}
\usepackage{balance}
\usepackage{booktabs}
\usepackage{bitpattern}
\usepackage{csquotes}
\usepackage{environ}
\usepackage[]{graphicx}
\usepackage[olditem,oldenum]{paralist}
\usepackage{pifont}
\usepackage[binary-units,group-separator={,}]{siunitx}
\usepackage{subcaption}
\usepackage{soul}  %
\usepackage{tikz}
\usepackage{xspace,relsize}
\usepackage{xstring}
\usepackage{blindtext}
\usepackage{multirow}
\usepackage{float}
\usepackage{url}
\usepackage[numbers]{natbib}

\usepackage{listings}

\definecolor{lstgreen}{rgb}{0,0.6,0}
\lstdefinestyle{plain}{%
	numbers=none,
	frame=none,
	xleftmargin=1pt,
	xrightmargin=1pt,
}

\lstdefinelanguage
[x64]{Assembler}     %
[x86masm]{Assembler} %
{morekeywords={CDQE,CQO,CMPSQ,CMPXCHG16B,JRCXZ,LODSQ,MOVSXD, %
			POPFQ,PUSHFQ,SCASQ,STOSQ,IRETQ,RDTSCP,SWAPGS,bextr,rorx,ud2,shrx}, %
	moreemph={rax,rdx,rcx,rbx,rsi,rdi,rsp,rbp, %
			r8,r8d,r8w,r8b,r9,r9d,r9w,r9b,
			eax, ebx, ecx, edx, esi, edi},
	morecomment=[l]{//},
}

\lstdefinelanguage{Rust}{%
	morekeywords={as,break,const,continue,crate,else,enum,extern,false,
			fn,for,if,impl,in,let,loop,match,mod,move,mut,pub,ref,return,Self,
			self,static,struct,super,trait,true,type,unsafe,use,where,while,
			abstract,alignof,become,box,do,final,macro,offsetof,override,priv,
			proc,pure,sizeof,typeof,unsized,virtual,yield},
	morekeywords=[2]{isize,usize,char,bool,str,String,u8,u16,u32,u64,u128,i8,i16,i32,i64,i128,f32,f64},
	sensitive=true,
	morecomment=[l]{//},
	morecomment=[l]{///}, %
	morestring=[b]{"},
}%

\usepackage{tikz}
\usetikzlibrary{shapes,arrows,positioning,shapes.multipart}

\newcommand{\circleone}{\ding{192}\xspace}

\newcommand{\circletwo}{\ding{193}\xspace}

\usepackage{caption}
\usepackage{paralist}
\renewcommand{\itemize}{\compactitem}
\renewcommand{\enumerate}{\compactenum}
\usepackage[font={small}]{caption}

\renewcommand{\paragraph}[1]{\medskip\noindent\textbf{#1}\hspace{1ex minus .1ex}}

\usepackage[disable]{todonotes}
\makeatletter%
\if@todonotes@disabled%
\else%
\fi
\makeatother

\newcommand{\version}[1]{#1}
\newcommand{\Rplus}{\protect\hspace{-.1em}\protect\raisebox{.35ex}{\smaller{\smaller\textbf{+}}}}
\newcommand{\Cpp}{\mbox{C\Rplus\Rplus}\xspace}

\usepackage{amsmath,amssymb,amsfonts}
\usepackage{algorithmic}

\usepackage{textcomp}
\usepackage{bmpsize}
\usepackage{xcolor}
\usepackage{lipsum}
\usepackage[colorlinks=true,urlcolor=black]{hyperref}
\usepackage{cleveref}

\newcommand{\toolname}{xTag\xspace}
\newcommand{\allocname}{mtmalloc\xspace}

\newcommand{\uaf}{use-after-free\xspace}

\newcommand{\runtimeOverhead}{\SI{24.9}{\percent}\xspace}

\newcommand{\relativeImprovementMarkUs}{\SI{25.2}{\percent}\xspace}

\newcommand{\artifacturl}{{\url{https://github.com/rub-syssec/xTag}}\xspace}

\def\BibTeX{{\rm B\kern-.05em{\sc i\kern-.025em b}\kern-.08em
		T\kern-.1667em\lower.7ex\hbox{E}\kern-.125emX}}
\begin{document}

	\title{\toolname: Mitigating Use-After-Free Vulnerabilities \\ via Software-Based Pointer Tagging on Intel x86-64}

    \author{\IEEEauthorblockN{Lukas Bernhard\IEEEauthorrefmark{1}, Michael Rodler\IEEEauthorrefmark{2}, Thorsten Holz\IEEEauthorrefmark{3}, and Lucas Davi\IEEEauthorrefmark{2}
        \IEEEauthorblockA{ \IEEEauthorrefmark{1}Ruhr University Bochum, Email:\{lukas.bernhard\}@rub.de  \\  \IEEEauthorrefmark{2}University of Duisburg-Essen, Email: \{michael.rodler, lucas.davi\}@uni-due.de  \\  \IEEEauthorrefmark{3} CISPA Helmholtz Center for
            Information Security, Email: \{holz\}@cispa.de }}
	}

	\maketitle

\begin{abstract}
	Memory safety in complex applications implemented in unsafe programming languages such as C/\Cpp is still an unresolved problem in practice. Such applications were often developed in an ad-hoc, security-ignorant fashion, and thus they contain many types of security issues.
    Many different types of defenses have been proposed in the past to mitigate these problems, some of which are even widely used in practice.
    However, advanced attacks are still able to circumvent these defenses, and the arms race is not (yet) over. On the defensive side, the most promising next step is a tighter integration of the hardware and software level: modern mitigation techniques are either accelerated using hardware extensions or implemented in the hardware by extensions of the instruction set architecture (ISA).
	In particular, \emph{memory tagging}, as proposed by ARM or SPARC, promises to solve many issues for practical memory safety. Unfortunately, Intel x86-64, which represents the most important ISA for both the desktop and server domain, lacks support for hardware-accelerated memory tagging, so memory tagging is not considered practical for this platform.

	In this paper, we present the design and implementation of an efficient, software-only pointer tagging scheme for Intel x86-64 based on a novel metadata embedding scheme. The basic idea is to alias multiple virtual pages to one physical page so that we can efficiently embed tag bits into a pointer.
	Furthermore, we introduce several optimizations that significantly reduce the performance impact of this approach to memory tagging.
	Based on this scheme, we propose a novel use-after-free mitigation scheme, called \toolname, that offers better performance and strong security properties compared to state-of-the-art methods. We also show how double-free vulnerabilities can be mitigated. Our approach is highly compatible, allowing pointers to be passed back and forth between instrumented and non-instrumented code \emph{without} losing metadata, and it is even compatible with inline assembly. We conclude that building exploit mitigation mechanisms on top of our memory tagging scheme is feasible on Intel x86-64, as demonstrated by the effective prevention of \uaf bugs in the Firefox web browser.
\end{abstract}


\section{Introduction}

Even after more than 20 years of research on memory corruption vulnerabilities in applications implemented in unsafe languages such as C/\Cpp, it is still an open problem how to efficiently and effectively mitigate these attacks in complex software systems.
While memory-safe systems programming languages, such as Rust or Go, are starting to become more popular, there is still a large number of legacy codebases written in unsafe C/\Cpp, such as web browsers, high-performance server applications, and operating system kernels.
Therefore, developing innovative and practical techniques to protect legacy C/\Cpp code bases against memory corruption attacks remains a crucial problem in practice.

Fortunately, due to the efforts of both academia and industry, it has become much more difficult to exploit memory errors in practice.
Multiple exploit mitigation techniques have been developed and deployed to production systems, such as non-executable memory, stack canaries~\cite{cowan1999stackcanaries}, address-space layout randomization (ASLR), and control-flow integrity (CFI)~\cite{Abadi2009cfi,ms-cfg,clang-cfi}.
However, these mitigations have forced attackers to resort to more advanced code-reuse attacks or data-oriented attacks~\cite{schuster2015coop,Hu2016dop,Ispoglou2018bop,Pewny2019steroids}.
As a result, the focus of mitigation research has shifted to identifying and mitigating the root cause of memory corruption so that entire classes of attack vectors can be prevented.
For example, various projects now use hardened heap allocators to reduce the likelihood of successful heap attacks~\cite{scudo,chromepartitionalloc}.
Although promising, software-based mitigation schemes are typically considered too slow for production use.
For example, Softbound+CETS, one of the strongest software-based memory safety solutions for C, adds more than \SI{100}{\percent} runtime overhead~\cite{nagarakatte2009softbound,Nagarakatte2010cets}.
As a result, many mitigation techniques are nowadays accelerated using hardware extensions or even fully implemented in hardware via extensions of the instruction set architecture (ISA).
For example, Intel is working on a hardware-based shadow stack called \emph{Control-flow Enforcement Technology} (CET)~\cite{shanbhogue2019security} and the ARM v8.3 architecture introduces cryptographic pointer authentication instructions~\cite{qualcomm-pa-whitepaper}.

One promising approach to improve memory safety are general memory/pointer tagging instructions, which can then be used to build exploit mitigation schemes.
Examples include ARM's \emph{Top-byte Ignore} (TBI) which allows software to use up to the eight most significant bits of a 64-bit pointer as a tag, and ARM's \emph{Memory Tagging Extensions} (MTE), which allows 4-bit tags to be assigned to each memory allocation.
While TBI is already available for 64-bit code in all Armv8 AArch64 processors, MTE will only be available in the future when processors with support for the ARM v8.5 architecture will become available.
Another example is SPARC's \emph{Application Data Integrity} (ADI)~\cite{serebryany2018mtag,sparcadi}, which also supports 4-bit tags.
Recently, Intel announced \emph{Linear Address Masking} (LAM)~\cite{intel-future}, a similar feature to ARM's \emph{Top-Byte-Ignore}, where address bits that are unused during address translation are allowed to have arbitrary values instead of being forced to be zero.
Unfortunately, on platforms that do not support such hardware extensions, it is still an open question how to efficiently implement these defenses.
This is particularly relevant for the widely used Intel x86-64 architecture, for which no CPUs with hardware-tagging support are available at the moment, thereby leaving the majority of desktop and server systems unprotected.

In this paper, we address this open challenge and show that hardware extensions, such as ARM's top-byte ignore feature, are actually \emph{not} required to implement efficient mitigation techniques based on the idea of pointer tagging. More specifically, we demonstrate that a purely software-based pointer tagging solution can be efficiently implemented on contemporary Intel x86-64 processors. Instead of relying on specialized---and often not yet available---hardware features, we show how the widely available memory management unit (MMU) can be repurposed to alias multiple virtual pages to one physical page, ensuring that the same allocation is referenced regardless of the embedded metadata.
Note that although we implement and evaluate software-based memory tagging on Intel x86-64, the same approach can also be implemented on other architectures with paging-based virtual memory.

Based on our software-based pointer tagging scheme, we propose a practical \uaf\ mitigation scheme, called \toolname, for the Intel x86-64 ISA.
We focus on \uaf\ vulnerabilities because they represent one of the most important bug classes in practice. A recent analysis by Google Project Zero~\cite{wild0dayp0} showed that since 2018, out of 45 zero-day vulnerabilities that were used ``in the wild'', 14 are due to \uaf\ bugs.
The Chromium team reports that \SI{36}{\percent} of all memory safety issues are due to \uaf~\cite{chromememsafety}.
Microsoft reports similar numbers for a period from 2015 to 2019: 557 CVEs, or roughly \SI{26}{\percent} of memory safety vulnerabilities at Microsoft, are due to \uaf~\cite{Bialek2020mtagsecurityanalysis}.
As such, we believe that an efficient mitigation will significantly improve software security.
Our mitigation scheme is based on the idea of embedding a 4-bit tag into heap pointers, the same size used by ARM MTE and SPARC ADI.
On each memory access, we can verify that the tag embedded into the pointer matches the tag stored in a disjoint shadow memory region.

Note that embedding metadata into pointers has a long tradition, ranging from fat pointers for bounds checking~\cite{nagarakatte2009softbound,akritidis2009baggyboundschecking} over low-fat pointers~\cite{Kwon2013lowfatpointers,Duck2016LowFat,Duck2017StackBP} to embedding cryptographic MACs~\cite{qualcomm-pa-whitepaper,ccfi2015}. Our approach is highly compatible (e.g., it even supports inline assembly), it allows pointers to be passed back and forth between instrumented and non-instrumented code \emph{without} losing metadata, and it induces a lower performance impact compared to existing methods~\cite{markus2020}.
In contrast to low-fat pointer techniques, our support for metadata invalidation enables mitigation of spatial memory safety vulnerabilities such as \uaf; we also show how double-free vulnerabilities can be efficiently detected.
For our defense, we propose several optimizations that significantly reduce TLB pressure, remove redundant checks, and reduce memory overhead.

To demonstrate the practical viability of the proposed approach, we implement a prototype based on the \emph{mimalloc} allocator and the \emph{LLVM} compiler framework. We modified the allocator to support our pointer metadata embedding scheme. For the \uaf detection scheme, we implemented an instrumentation pass in the LLVM compiler framework. Our instrumentation utilizes the pointer tagging scheme to retrieve metadata about the current state of the reference object. Furthermore, we develop several optimization techniques on the compiler level to reduce the overhead induced by instrumentation.
To demonstrate the effectiveness of our approach, we instrument the Firefox web browser with our defense and show in several case studies that we can successfully detect and prevent different bugs in such a complex, real-world application.
Furthermore, we show that our approach is more efficient in terms of performance overhead compared to state-of-the-art approaches in this area. On the SPEC CPU2017 benchmark, we achieve a significant reduction in geomean runtime overhead.
In particular, we reduce the relative overhead for \uaf mitigation by \relativeImprovementMarkUs compared to the fastest competing approach.
An extensive evaluation of mitigation configurations, including different tag sizes and selection strategies, highlights the impact on TLB pressure and the resulting performance impact.
The evaluation shows that our mitigation is well suited for high-throughput scenarios (i.e., allocation- and memory-intensive scenarios).

\paragraph{Contributions}
To summarize, our contributions are: %
\begin{enumerate}
	\item We present the first efficient and practical scheme for embedding an \emph{invalidatable} metadata tag inside of pointers using aliased page mappings. This approach can be implemented on platforms \emph{without} any kind of hardware support, most importantly Intel x86-64, and is even compatible to non-instrumented code such as inline assembly.
	\item Based on this scheme and several optimizations, we present a \uaf mitigation scheme based on validating the pointer tag. Our prototype implementation can successfully prevent \uaf vulnerabilities in the Firefox web browser. In the SPEC CPU2017 benchmark suite, we achieve a geomean runtime overhead of \runtimeOverhead, a reduction of relative overhead by \relativeImprovementMarkUs compared to the state-of-the-art approach.
\end{enumerate}

\paragraph{Code Availability}
To enable reproducibility of our results, the code of \toolname is available at \artifacturl under an open-source license.

\section{Technical Background}%
\label{sec:background}

We first review the basics of memory tagging, low-fat pointers, and \uaf vulnerabilities given that these are important concepts that are necessary for understanding the rest of this paper.

\subsection{Memory Tagging and Low-Fat Pointers}
\label{sec:mtagbackground}

Tagged memory has diverse applications for building secure systems, mitigations, and sanitizers~\cite{Watson2015cheri,song2019soksanitizing}.
Memory tagging~\cite{mtpatent} can be used to facilitate a lock-and-key mechanism that protects against temporal memory safety violation as well as a subset of spatial violations.
The typical approach is as follows: Memory ranges are assigned locks at the time of allocation. All mechanisms to access memory are modified such that the correct key must be provided alongside the pointer. Generally speaking, the lock and key share the same value, which is referred to as a \emph{tag}. Any failure to provide the correct tag triggers an exception, usually leading to program termination.

This scheme can protect against spatial security violations when the tag of an out-of-bounds pointer does not match the tag required to access the memory location. For example, two adjacent arrays \emph{a1} and \emph{a2} are assigned distinct tags \emph{t1} and \emph{t2}, respectively. Due to an out-of-bounds error, \emph{a1} is indexed such that the address corresponds to \emph{a2}. As the memory access provides \emph{t1} instead of \emph{t2}, the violation can be successfully detected.
However, not all spatial safety violations are mitigated: an array within a struct might overflow into an adjacent field. Since all fields in the struct are part of the same allocation, they share the same tag and do not violate the memory tagging policy.
Temporal safety violations are mitigated by assigning a new tag when memory is freed. Suppose \emph{a1} is freed, which implies that the memory region's tag is changed from \emph{t1} to \emph{t1'}. All pointers to \emph{a1} currently held in local variables or stored in memory are invalided as they are still bound to the tag \emph{t1}.

Hardware implementations of memory tagging exist on only a few architectures, i.e., they are already available on SPARC via ADI and will be available in the future in the ARM 8.5 architecture through the combination of ARM's Top-byte Ignore (TBI) and Memory Tagging Extensions (MET). Retrofitting memory tagging to established architectures poses a significant challenge as the corresponding tag must be stored alongside each pointer. Modifying the pointer size breaks the ABI and requires significant changes to handwritten assembly code. Instead, the SPARC and ARM architecture embed tags into the \emph{native} pointer representation. Both platforms dedicate a range of pointer bits to store the tag. This range is ignored by the MMU during address translation and is instead used to store tags.

Current Intel x86-64 processors neither offer a hardware implementation of memory tagging nor an address mode that ignores a range of pointer bits.
While the verification could be implemented in software, efficiently storing tags \emph{without} breaking the ABI (e.g., support for non-instrumented libraries and the system call interface) remains an open challenge that we address in this work. Furthermore, we also need to ensure backward compatibility, e.g., usage of inline assembly.

\emph{Low-fat pointers}~\cite{Kwon2013lowfatpointers,Duck2016LowFat,Duck2017StackBP} were proposed as an approach for encoding metadata bits into the machine representation of pointers.
Existing designs encode static information by placing allocations into specific heap regions (e.g., grouping all allocations of a specific size).
As low-fat pointers can be dereferenced natively, they achieve compatibility with uninstrumented code.
Extracting the heap region from a pointer allows to derive the allocation boundaries corresponding to a pointer.
Taking advantage of this metadata, existing mitigation schemes protect against out-of-bound memory accesses.
Unfortunately, none of the existing schemes allow for an invalidation capability, i.e., encoded metadata remains valid even \emph{after} freeing an allocation. Therefore, they are unable to mitigate spatial memory safety vulnerabilities such as \uaf attacks.

\subsection{Use-After-Free Vulnerabilities}
\label{sec:uafbackground}

The C/\Cpp programming languages leave most of the responsibilities for memory management to the programmer. Memory objects, whose lifetime is not bound to a stack frame, are typically allocated in the heap region. The management of the memory within the heap region is typically performed by a standard allocator library. For example, the C~standard library provides the \emph{malloc} and \emph{free} functions to allocate and deallocate memory objects, respectively. When to allocate and deallocate an object is in the hands of the programmer calling the allocator APIs. As a consequence, there are many types of errors related to memory management on the heap. Not releasing memory that is no longer referenced leads to inconvenient memory leaks. More critically, freeing a memory object twice---a so called \emph{double-free} vulnerability---can corrupt an allocator's internal state, potentially leading to further memory corruption.

Furthermore, one of the most critical bugs is the \uaf condition, where the program uses a dangling pointer to access an already freed object. Listing~\ref{lst:uafexample} shows an example of a \uaf\ bug: the two pointers \emph{p} and \emph{q} reference the same memory object. After the object is freed via the pointer \emph{p}, the \emph{q} pointer becomes dangling. To corrupt memory of the application, an attacker must ensure that another object is allocated at the same address as the previous allocation. In this case, the \emph{u} pointer references such an unrelated object. Using the dangling \emph{q} pointer leads to access to the \emph{u} object instead of the actual intended object.

\begin{lstlisting}[caption={Example of a \uaf\ bug (a temporal memory safety violation).}, label=lst:uafexample, float=t, belowskip=-1.2 \baselineskip]
uint32_t *p, *q;
char* u;
p = malloc(8); // allocates uint32_t[2]
q = p + 1;     // q references second uint32_t
// [...]
free(p);
u = malloc(8); // likely(u == p)
// [...]
*q = 21; // UAF bug: Access modifies unrelated
         // object u in memory leading to
         // memory corruption
\end{lstlisting}

\section{Design}%
\label{sec:design}

\begin{figure*}[t]
	\centering
	\includegraphics[width=0.7\linewidth]{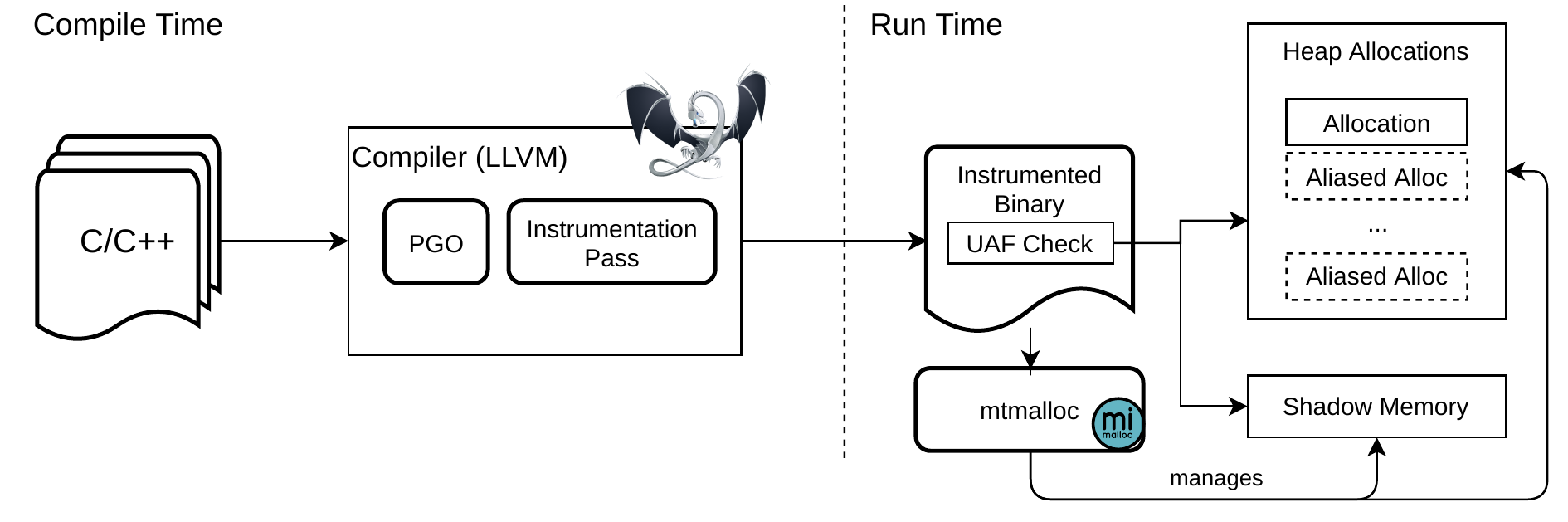}
	\caption{Schematic overview of \toolname's architecture.}%
	\label{fig:architecture}
\end{figure*}

In this section, we introduce our metadata embedding scheme for pointers on the Intel x86-64 ISA (i.e., on an ISA \emph{without} hardware support for pointer tagging).
We embed metadata in heap pointers, while maintaining compatibility with inline assembler code and other non-instrumented code.
Based on this design, we present an efficient UAF mitigation method.
Our UAF mitigation leverages the metadata embedding to provide a probabilistic defense against temporal safety violations, namely use-after-free of heap-allocated data.
\Cref{fig:architecture} shows the overall architecture of our system and the prototype implementation called \toolname.

Our system takes as input the source code (C/\Cpp) of an application.
In our prototype implementation, we use the LLVM compiler framework to insert \uaf\ checks into the compiled binary based on the metadata stored in the pointer and the shadow memory (see \Cref{sec:uafdetection}).
Note that \toolname requires access to the source code for inserting these \uaf checks, and that closed-source applications cannot be protected by parties other than the vendor.
We do not believe that this will hinder adoption, as this drawback also applies to many widely deployed mitigations, such as e.g., LLVM's CFI or stack cookies.
We use profile-guided optimizations (PGO) provided by the LLVM compiler framework to reduce branching in emitted \uaf\ checks (see \Cref{sec:opt}).
The instrumented binary is linked to a memory allocator, called \emph{\allocname}, in our system.
Our custom memory allocator manages both the shadow memory, where the memory tags are stored, and the heap allocations, which require multiple aliased virtual memory mappings due to metadata embedded into the pointer (see \Cref{sec:sub:metadata}).

\subsection{Metadata Embedding Scheme}%
\label{sec:sub:metadata}

We design a generic allocator \emph{\allocname} that supports an efficient software implementation of metadata embedding in pointers for the x86-64 ISA.
We then show how to build a UAF mitigation on top of this allocator.
Note that our embedding scheme is not limited to \uaf mitigations in its applications. Any defense mechanism that relies on a metadata embedding scheme (e.g., CSP~\cite{kuznetsov2014CPI} and further security applications~\cite{song2016HDFI}) is supported and can benefit from our software-based pointer tagging technique. As a proof-of-concept, we also demonstrate how double-free vulnerabilities can be mitigated. 
Broadly speaking, we reserve a predefined range of bits to store metadata inside a pointer.
From a CPU's perspective, the embedded metadata bits correspond to changes of the virtual address. Hence, to preserve compatibility with non-instrumented code, we need to ensure that pointers reference the same allocation \emph{regardless} of the embedded metadata bits.
As such, we propose a technique that maps distinct pointers to the same allocation.
To do so, we map the physical memory backing an allocation to \emph{multiple} virtual addresses.
In particular, distinct pointers map to the same physical memory if and only if the pointers differ only in the embedded metadata bits.
Consequently, the same physical memory corresponding to an allocation is referenced regardless of the embedded metadata bits.
At the same time, this ensures that non-instrumented code can still dereference the pointer.

\begin{figure}[t]
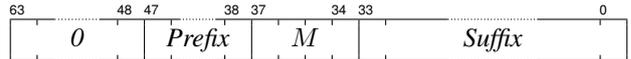

	\centering
	\bitpattern[numberFieldsTwice,littleEndian,numberBitsAbove,startBit=63]{0}[16][5]{Prefix}[10][4]{$M$}[4]{Suffix}[34][10]/

	\caption{Heap pointer format in our embedding scheme for a canonical \SI{64}{\bit} pointer with \SI{4}{\bit} embedded metadata (designated as $M$).}%
	\label{fig:heappointer}
\end{figure}

We propose a scheme that defines a format for heap pointers that reserves space for metadata bits.
While the general approach can be extended to stack pointers and global variables, our design focuses on heap pointers as we aim at constructing a \uaf\ defense focusing on heap-allocated objects (see \Cref{sec:uafdetection:attacker}).
\Cref{fig:heappointer} shows the format of heap pointers in an x86-64 address space.
The first part of the pointer is set to 0 to conform to canonical address requirements in current x86-64 implementations.
We use a pointer prefix to quickly distinguish between tagged heap pointers and other untagged pointers.
Then we use \SI{4}{\bit} for storing metadata inside of the pointer ($M$ in \Cref{fig:heappointer}).
We chose \SI{4}{\bit} since this tag size offers good performance characteristics (see our evaluation in \Cref{sec:evaluation}), while still offering good protection (see a discussion in \Cref{sec:securityanalysis}).
Furthermore, this is in line with hardware-backed tagging solutions like SPARC ADI and ARM MTE, which also use \SI{4}{\bit} tags.

As we focus on heap pointers, only a subset of pointers will be embedding metadata.
Hence, further instrumentation is required to distinguish between pointers with and without metadata.
We reserve a virtual memory range that spans $2^{38}$ bytes during initialization, so that no other memory mapping falls into the same range.
It is easy to recognize a heap pointer in our scheme because all heap allocations have the same fixed pointer prefix if and only if they are managed by \allocname.
Coexisting heap allocators might allocate memory areas with different prefixes, but \allocname\ enforces a common prefix for all heap allocations.

In our proposed scheme, bits 37 to 34 (i.e., 4~bits) are designated as \emph{metadata bits}.
Note that the number of metadata bits can be adjusted, and we evaluate in \Cref{sec:evaluation} the performance impact of this parameter.
By default, pointers returned by our allocator always have these metadata bits set to 0. %
The application is then free to use the metadata bits in the pointer, i.e., for implementing mitigation schemes.
This leaves bits 33 to 0 at the discretion of the allocation algorithm.
Upon an allocation request, the allocator searches for a free slice of virtual memory within this range.
The allocation algorithm is independent of our pointer format and is therefore not specified any further.

Efficiently implementing a pointer tagging scheme with aliased virtual page mappings is highly challenging for multiple reasons.
First, the page tables of the process grow as they need to store multiple entries.
Second, the increased number of system calls imposes additional context switch overhead.
Third, and most importantly, the CPU's caching mechanisms are impacted.
Primarily, the pressure is increased on the Translation Lookaside Buffer (TLB), which caches virtual to physical address translation.
In \Cref{sec:evaluation}, we present an evaluation of these effects.
In \Cref{sec:opt}, we describe how we tackle these challenges by introducing several new optimization techniques.

\begin{figure}[t]
	\centering
	\includegraphics[width=\linewidth]{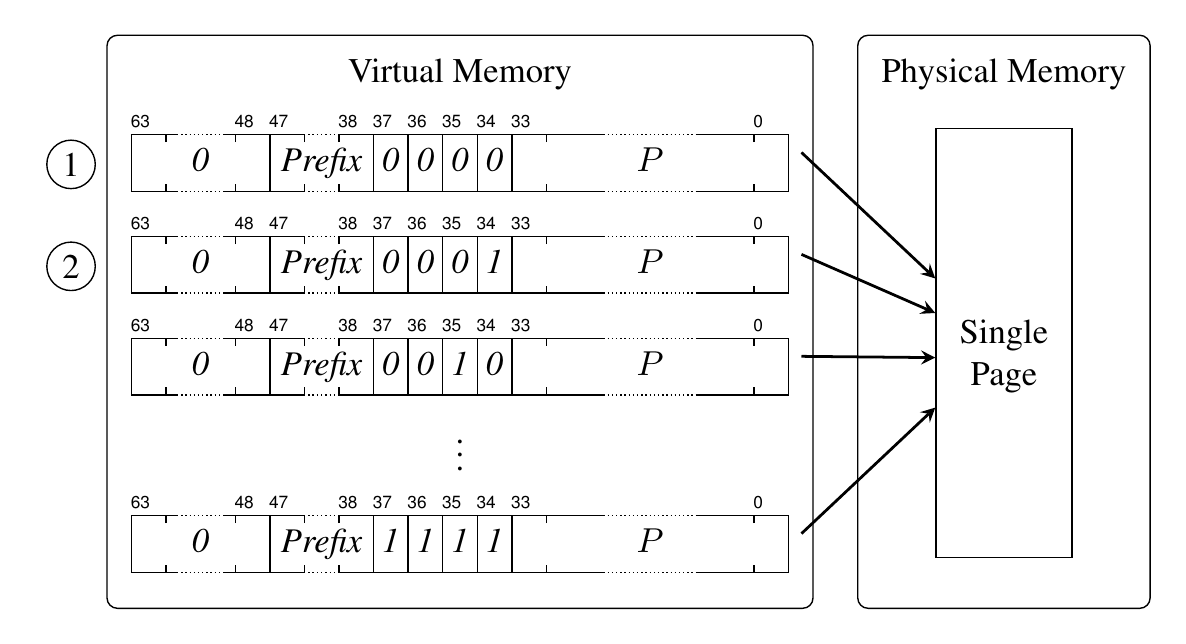}
	\caption{Various metadata embeddings that all map to the same physical page. The allocator maps virtual memory for all possible metadata embeddings to ensure every pointer can still be dereferenced, regardless of embedded metadata.}%
	\label{fig:metadataexample}
\end{figure}

To illustrate our approach, \Cref{fig:metadataexample} shows such an allocation with an empty tag of $0000$ at \circleone.
According to the defined pointer format, 4 bits are available for metadata usage.
Changing these bits influences the virtual address used during memory accesses.
As an example, the application might embed the bit pattern $0001$ into the metadata field, which results in the pointer depicted at \circletwo\ in \Cref{fig:metadataexample}.
As memory accesses via this pointer must resolve to the same allocation, we have to ensure that the virtual address maps to the same physical address.
This property is constructed by mapping the virtual page at \circleone\ to the same physical address as virtual page at \circletwo.
As a result, the two addresses can be used equivalently, facilitating an arbitrary bit pattern in the 4 metadata bits.
As a consequence, \allocname\ needs to map the physical address $2^4 = 16$ times.
This induces additional overhead by requiring more system calls and larger page tables, and decreases the effectiveness of the CPU's caching mechanisms.
However, our experiments show that most of these effects are either negligible (see \Cref{sec:evaluation}) or can be reduced with new optimizations (see \Cref{sec:opt}).

Note that the presence of embedded metadata is fully transparent to the application code.
Due to the aliased virtual memory mappings, pointers with any embedded metadata are also regular pointers and can be dereferenced.
This is a significant and important advantage when dealing with code that is not instrumented.
In prior work, pointers containing metadata had to be converted to regular pointers and back at \emph{each} boundary to non-instrumented code.
This includes the usage of inline assembly, non-instrumented libraries, or the system call interface.
In contrast, our metadata scheme is highly compatible, allowing pointers to be passed back and forth between instrumented and non-instrumented code \emph{without} losing the metadata.
This allows our scheme to work with complex targets, such as web browsers, that use multiple source languages, complex runtimes, legacy code, and are highly optimized.

\subsection{Use-after-free Mitigation Scheme}%
\label{sec:uafdetection}

Based on our software-based pointer tagging method, we introduce a use-after-free mitigation scheme that uses the metadata bits to mitigate this type of vulnerabilities.

\paragraph{Threat Model}%
\label{sec:uafdetection:attacker}
In the remainder of this paper, we assume the following threat model for our mitigation scheme: the adversary knows of at least one use-after-free vulnerability in a given vulnerable program.
This vulnerability allows the adversary to force the program to read from or write to a dangling pointer; this operation accesses a deallocated memory object that potentially overlaps with one (or more) other memory objects.
Use-after-free vulnerabilities represent a temporal memory safety violation that is most commonly associated with heap-allocated data.
While it is possible to encounter such conditions with stack-allocated data (i.e., use-after-return), current data~\cite{wild0dayp0,Bialek2020mtagsecurityanalysis} suggests that the vast majority of known use-after-free vulnerabilities are heap-based.
Furthermore, to cover use-after-return, the compiler can simply transform stack allocations to heap allocations to be covered by \toolname's protection.
As such, we focus solely on heap-based use-after-free vulnerabilities in this work.

We also assume that orthogonal defenses against spatial memory safety issues are in place~\cite{akritidis2009baggyboundschecking,Kwon2013lowfatpointers,Kroes2017midfat,Duck2016LowFat}.
However, to bypass \toolname's protection using a spatial memory error, the attacker would need an arbitrary write primitive to corrupt \toolname's shadow memory.
In this case, bypassing \toolname is probably not necessary anymore, as the attacker can essentially corrupt the whole address space already.

\paragraph{Mitigation Scheme}%
\label{sec:uafdetection:design}
Building on the metadata embedding scheme, we design a probabilistic exploit mitigation targeting \uaf\ bugs.
Similar to prior work on \uaf\ detection, we utilize a lock-and-key-style mechanism~\cite{Nagarakatte2010cets} to detect \uaf\ conditions.
More specifically, we assign a randomly chosen 4-bit value (called \emph{tag} further on) to every heap allocation.
This tag is embedded in the pointer returned by the allocation function and is additionally stored in a shadow memory region.
When memory is accessed via a pointer, the embedded tag is compared to the corresponding tag stored in the shadow memory region.
Any mismatch is considered a \uaf\ and terminates program execution.

Prior work on \uaf\ mitigations has used a similar type of mitigation~\cite{Nagarakatte2010cets,oscar2017}.
However, these earlier approaches have difficulties to store the reference value.
Typically, fat pointers are used, which break compatibility with non-instrumented code, such as inline assembly.
Because of this incompatibility, they have not been widely used in practice.
Low-fat pointer schemes~\cite{Kwon2013lowfatpointers,Duck2016LowFat,Duck2017StackBP} improve compatibility, but do not allow neither updating nor invalidating of embedded metadata. As further explained in \Cref{sec:relatedWork}, it is precisely this limitation that precludes a \uaf mitigation scheme base on low-fat pointers.
In contrast, our pointer embedding scheme allows us to handle scenarios where a pointer is passed to non-instrumented code, which returns the pointer back to instrumented code.
In this case, the returned pointer still contains the same embedded metadata tag, and the instrumented code can again check for \uaf.
As a result, our \uaf\ mitigation scheme can be enabled in many software projects, even when various programming languages, runtimes, and inline assembly code are used.
In this respect, our mitigation scheme is similar to \uaf\ mitigations that were proposed for the ARM architecture with hardware support for memory tagging~\cite{serebryany2018mtag}.
Note that the Intel x86-64 ISA does not support such methods. %
In contrast, our proposed scheme can be applied to complex code bases on commodity hardware.
For example, we were able to successfully instrument the \Cpp part of the Firefox web browser.

For each allocation, we generate a tag that is embedded in the pointer returned by allocation functions (e.g. \emph{malloc} and \emph{new}).
Additionally, we assign the same tag to the memory range corresponding to the allocation.
The memory tags are stored disjointly inside a shadow memory region.
The granularity of our heap tagging is 16 bytes, i.e., 16 bytes of heap memory map to 1 byte of shadow memory.
This implies a minimum allocation granularity of 16 bytes, otherwise memory ranges corresponding to distinct allocation might share the same shadow memory entry.
While this is not the most memory-efficient allocation granularity, our allocator does not impose unreasonable overheads.
First, typical heap allocations are larger than 16 bytes (the size of two pointers).
Second, high-performance heap allocators already align memory to 8 or 16 byte granularity to better utilize caching mechanisms of the CPU.
In Section~\ref{sec:evaluation}, we analyze the performance of our scheme in detail.

We chose the shadow memory scheme to efficiently implement the runtime checks.
Every pointer into an allocation has a trivial linear mapping to shadow memory.
This means that during tag lookup, there is no need to find the base address of the allocation via an expensive traversal operation.
The downside of this shadow memory scheme is that memory usage of the shadow memory grows with the number of bytes allocated.
As such, the memory overhead of the shadow memory amounts to roughly \SI{6}{\percent} for all heap allocations.
This is in line with other works that utilize memory tagging~\cite{serebryany2018mtag} and is lower than the memory overhead induced by other recent \uaf\ defenses (e.g., \SI{15}{\percent} in MarkUs~\cite{markus2020} and \SI{61.5}{\percent} in Oscar~\cite{oscar2017}).
We present a full measurement of the memory overhead in \Cref{sec:evaluation}.

We instrument all pointer usages to check whether the tag embedded in the pointer and the tag stored in shadow memory have the same value.
Listing~\ref{lst:pseudocinstrumentation} shows our pointer usage instrumentation in pseudo-C code.
Since pointers to stack and globals are not tagged at all, we can skip further checks on these pointers.
This is possible by verifying that the pointer starts with the heap prefix, as shown in line 1 in Listing~\ref{lst:pseudocinstrumentation}.
For heap pointers containing embedded metadata, extraction of the tags boils down to a shift operation followed by a logical \emph{and} operation (line 2).
The lookup of a tag in shadow memory starts with computing the offset for the heap base.
As there is a linear mapping between heap and shadow memory, we can simply use the offset to the heap base to perform the lookup into shadow memory.
This offset is divided by 16 to account for the tagging granularity.

\begin{lstlisting}[caption={Pseudo-C that shows the memory tag lookup and reporting.}, label=lst:pseudocinstrumentation, float=t, belowskip=-1.2 \baselineskip]
// verify we have a heap pointer
if ((pointer ^ heap_prefix) >> 38 == 0) {
  // extract tag from pointer
  tag = (pointer >> 34) & 0xf;
  // retrieve tag byte from shadow memory
  heap_offset = pointer & 0x3ffffffff;
  shadow_tag = shadow_memory[heap_offset / 16];
  if (tag != shadow_tag) { // validate tag
    __report_uaf();
  }
}
\end{lstlisting}

\paragraph{Extension: Double-Free Detection}
Double-free bugs are a related temporal memory safety issue where the attacker corrupts the allocators internal state by abusing a deallocation of an already deallocated memory object.
More specifically, the free list of the allocator contains the same pointer twice, which leads to the allocator returning the same address for different calls to \emph{malloc} or \emph{new}.
Similar to \uaf bugs, there are now two semantically different pointers that refer to the same memory location.
However, in contrast to \uaf, such double-free bugs can be mitigated in the allocator itself (i.e., no instrumentation is necessary).

We extend \toolname, more specifically our allocator \allocname, to detect and prevent \emph{double-free} bugs using memory tags.
Once an allocation is returned to the allocator (e.g., \emph{free} or \emph{delete}), \allocname performs an additional check that verifies that the pointer-embedded tag matches the tag stored in shadow memory.
A tag mismatch indicates a double-free issue, hence we terminate execution.
To ensure that subsequent attempts to free the allocation again are detected deterministically, we override the first corresponding tag in shadow memory.
This is sufficient to detect double-free bugs because a call to \emph{free} receives a pointer to the beginning of the allocation as an argument.
Enabling the double-free mitigation did not have any measurable runtime or memory overhead compared to the \toolname version without the double-free mitigation.

\subsection{Design Optimizations}%
\label{sec:opt}

Allocator modifications and runtime instrumentation both degrade the performance of the system, optimizing either aspect reduces the overall overhead.
The allocator induced overhead stems primarily from two effects.
First, mapping memory multiple times increases the number of system calls to establish the mapping in the first place.
Second, the pressure on the translation lookaside buffer (TLB), a memory cache that is used to reduce the time taken to access a memory location, increases and this can significantly affect the performance.

Our instrumentation inserts runtime verification checks which consume both computing cycles and memory bandwidth.
A single check consists of only a short sequence of assembly instructions (see Listing~\ref{lst:asmpointervalidation} on page~\pageref{lst:asmpointervalidation}).
As such, the primary optimization possibility is to reduce the number of runtime checks.
Hence, we introduce several methods that leverage information provided by the compiler to reduce the number of necessary runtime checks while instrumenting the target program.

\subsubsection{Reducing TLB Pressure}
\label{sec:tlboptimization}

\Cref{sec:uafdetection} introduced a random selection of tags for the sake of simplifying the description.
Hence, allocations adjacent on a physical page tend to be placed on distinct virtual pages.
Accessing a single allocation establishes a TLB entry caching the translation from the virtual page to the corresponding physical page.
However, the cache entry is unlikely to speed-up access to an allocation belonging to the same physical memory as the virtual page differs with high likelihood due to different embedded metadata.
Therefore, iterating over multiple allocations has a high probability to trigger multiple page-walks in order to establish the TLB entries, even if the allocations are placed right next to each other in physical memory.
As a result, the increased number of page-walks slows down execution.
In addition, the TLB cache is limited in size and establishing a new entry might evict an entry needed in the near future.
As shown in our performance evaluation in \Cref{sec:evaluation}, the increased TLB pressure results in a workload-dependent slowdown.

We reduce the TLB pressure with two orthogonal optimizations. First, we implement huge page support.
Huge pages increase the size of a virtual-to-physical page mapping from \SI{4}{\kilo\byte} to \SI{2}{\mega\byte}. As a result, the same number of TLB entries cache the translation of a larger span of virtual addresses to physical addresses.
Second, the assignment of tags to allocations can be optimized in order to reduce TLB pressure, which in turn improves performance.
The key idea here is to keep assigning the same tag to allocations on the same page as long as memory on this very page is not reused.
When the allocator selects a new page for fulfilling allocations, a random tag is generated for the page and all allocations from this page are tagged with the same value.
Hence, \emph{initially} all allocations share the same virtual page and therefore require just a single TLB entry.
However, to not compromise the security properties of our mitigation scheme, memory returned to the allocator cannot simply reuse the same tag.
Otherwise, the optimization would become susceptible to \emph{Heap Feng Shui}~\cite{sotirov2007heap}.
Instead, the allocator tracks allocations released by the application in a separate set.
Only if a virtual page is filled with allocations and can no longer serve new allocations, we randomly generate a new fresh tag for this page.
The separately tracked set of freed allocations is now made available to new allocations again.
All further memory allocations receive the newly assigned tag.
Thus, memory reused by the allocator always contains a random tag, while still aiming to reduce the diversity of tags for better TLB caching.

Note that the runtime validation of tags always compares the tag embedded in the pointer with the tag stored in shadow memory.
Even if the allocator re-randomized the tag for new allocations within a page, the currently existing pointers and allocations remain valid.
We refer to this optimization as \emph{generational tag assignment}.
The positive effect of this optimization becomes apparent during the evaluation in \Cref{sec:evaluation}.

\subsection{Optimizing Instrumentation}
\label{sec:optimizing}

When considering temporal safety violations, we can often deduce that certain program paths do not affect temporal safety properties.
For example, a program path that does not free memory cannot invalidate \uaf\ checks.
As such, we can often remove redundant checks in our instrumentation.
A discussion of this optimization's security implications is deferred to \Cref{sec:securityanalysis:multithreading}.

\paragraph{Repeated Accesses via Same Pointer}
Multiple memory accesses via the same pointer are validated just once if the allocation cannot be freed in between.
Consider the code snippet in Listing~\ref{lst:pseudocmultipleaccess}.
The code starts with dereferencing pointer \texttt{x}.
Shortly afterwards, the very same pointer is dereferenced again.
The compiler must emit a second load, as the interleaving write via \texttt{y} might change \texttt{*x}.
Assuming our runtime validation verifies \texttt{x} during the first load, the validity must hold for the second load.
We derive this property by analyzing code paths between memory accesses via the same pointer.
Under the condition that none of the paths between two accesses might free memory, a later check can be omitted.
This optimization is applicable only if the former access dominates the latter, i.e., we can prove statically that the former access must have been executed before the latter memory access.
Otherwise, a path might reach the latter access without passing through a runtime verification.
Again, discussion of this optimization's security implications is deferred to \Cref{sec:securityanalysis:multithreading}.

\begin{lstlisting}[caption={Pseudo-C showing multiple accesses via the same pointer.}, label=lst:pseudocmultipleaccess, float=t]
int f(int* x, int* y) {
  int result = *x;
  // might override *x if pointers alias
  *y = 0;
  // must re-read *x due to previous write
  result += *x;
  return result;
}
\end{lstlisting}

\paragraph{Partially-Aliasing Pointers}
Two pointers referring to the same allocation at distinct offsets into the allocation are called \emph{partially-aliasing}.
E.g., pointers to different fields of a struct refer to same allocation but point to distinct addresses.
Because allocations cannot be partially freed, validating a single pointer implicitly validates all partially-aliasing pointers as well precisely because they point into the very same allocation.
Consequently, we can omit checking a memory accesses via pointer $P$ if
\begin{inparaenum}[(1)]
	\item a partially-aliasing pointer $P_a$ is verified in a code path dominating the access in question, and
	\item no memory is freed between the validation of $P_a$ and the memory access via $P$.
\end{inparaenum}

As an example, consider the code in Listing~\ref{lst:pseudocaliaspointers}.
Setting the two member variables \texttt{x} and \texttt{y} accesses memory via two distinct pointers.
The pointers partially alias each other as they must belong to the same allocation.
In addition, the first memory access dominates the second (i.e., must happen before) and memory cannot be freed between the accesses.
This implies that verifying the first pointer implicitly ensures that the second pointer is not dangling.

\begin{lstlisting}[caption={Pseudo-C showing two memory writes. The accessed memory locations are partially-aliasing as they must belong to the same allocation.}, label=lst:pseudocaliaspointers, float=t, belowskip=-1.2 \baselineskip]
typedef struct { int x; int y; } S;
S* s = malloc(sizeof(S));
s->x = 0;
s->y = 1;
\end{lstlisting}

\paragraph{Validating in the Caller}
When considering small helper functions that do not free memory, we can sometimes move runtime checks out of the function.
For example, consider a simple comparison function between two memory objects such as an implementation of the \Cpp equality operator.
This small helper function does not change temporal memory safety properties and only accesses two memory objects.
We can now introduce the assumption that all pointers passed to this function are already safe and completely eliminate the checks from the function.
However, now any caller of the function must perform the \uaf\ checks on behalf of the callee.
At first, this does not reduce the number of necessary checks, but if we consider passing a pointer to a callee function as a pointer access, we can often eliminate these checks as part of our \emph{repeated access via same pointer} optimization.

\section{Implementation}%
\label{sec:implementation}

Our prototype implementation of \toolname\ consists of two major components:
\begin{compactenum}
	\item \emph{\allocname}, a memory allocator managing the memory tags and aliased allocations and
	\item a LLVM compiler pass that leverages profile-guided optimization (PGO) to efficiently instrument the code.
\end{compactenum}

We implemented the proposed allocator design on top of \emph{mimalloc} \version{v1.6.1}, a high-quality and high-performance allocator developed at Microsoft~\cite{mimalloc}.
While the original mimalloc supports all major operating systems, i.e., Windows, Linux, and OS X, our modifications target the Linux kernel to demonstrate the feasibility of the approach.
Our prototype relies on the Linux kernel memory management API to map physical memory multiple times with the mmap flag \texttt{MAP\_SHARED}.
This flag is intended to facilitate user-mode shared memory between multiple processes, both reading and writing to the same physical memory.
However, it also allows mapping the same physical memory multiple times into a single process.
Due to mimalloc's support for Windows and OS X, we expect modest porting efforts to support other operating systems.
Both of these operating systems support mapping physical memory multiple times either via \texttt{MAP\_SHARED} on OS X or \texttt{MapViewOfFileEx} on Windows.

Our allocator exports two global variables, \texttt{heap\_start} and \texttt{shadow\_base}.
The symbol \texttt{heap\_start} provides the start of a \SI{16}{\giga\byte} memory region which holds all dynamic memory allocations.
This allows our instrumentation to distinguish whether a specific pointer is within the heap region.
Note that the current heap size of \SI{16}{\giga\byte} is an implementation detail and can be increased to support memory-intensive applications.
Differentiation is necessary as non-heap pointers lack a tag and consequently they are exempt from runtime validation.
Mapping heap pointers to the corresponding shadow memory region mandates the base address of said region, provided from the allocator via the symbol \texttt{shadow\_base}.
Runtime instrumentation relies on the two symbols, while verifying if the embedded tag matches the reference value in shadow memory.
The instrumentation is emitted by our link-time optimization (LTO) pass for LLVM version~10 consisting of approximately 3k lines of code.
We release both our allocator and instrumentation passes such that our results can be reproduced.
The repository is available at \artifacturl.

\paragraph{Assembly Sequence for Validation}
As our mitigation needs to check a significant amount of pointers at runtime, efficient implementation is vital for a small performance penalty.
While Listing~\ref{lst:pseudocinstrumentation} in Section~\ref{sec:uafdetection} shows pseudo-code for validation, Listing~\ref{lst:asmpointervalidation} shows the actual assembly code emitted by our compiler pass.
We start with a \texttt{rorx} instruction which rotates the bits in \texttt{ptr} such that the tag is stored in \texttt{cl}.
Furthermore, the 30 leftmost bits of \texttt{rcx} now hold the offset into shadow memory.
We commence with extracting this offset into \texttt{rax} with a \texttt{shrx} instruction.
Note that \texttt{shadow\_base} is a global variable, exported by our allocator.
It holds the shadow memory base address.
In addition, this base address is aligned such that the lowest byte is precisely the constant we need for \texttt{shrx} such that the offset is stored in \texttt{rax}.
The subsequent \texttt{cmp} instruction compares the extracted tag in \texttt{cl} with the counterpart in shadow memory.
Assuming the tags match, we resume regular program execution.
Any mismatch not yet substantiates a usage of a dangling pointer.
Instead, we have to fallback on a check whether \texttt{ptr} is a heap pointer in the first place.
This check is implemented by comparing the prefix bits of \texttt{ptr} with the global variable \texttt{heap\_start}, exported by the allocator.
A different prefix means the pointer is not managed by our allocator (e.g., located on the stack or in the .bss section).
We allow program execution to proceed since we do not need to check this pointer.
Leveraging PGO information, we can invert the branch condition if training data suggests a single check mostly validates non-heap pointers.
In this case, a prefix comparison proceeds the
tag validation which is moved to the fallback branch.

Assume a specific code location mostly validates non-heap pointers.
Failing tag validation (and subsequent fallback code) is unnecessarily expensive.
Instead, reordering tag validation and fallback code is more efficient.
As determining the better ordering at compile-time is difficult, we leverage profile-driven feedback instead.
Assuming the profiling workload is representative, we pick a code sequence minimizing runtime overhead in production.

\lstset{language=[x64]Assembler}
\begin{lstlisting}[caption={x86-64 assembly sequence runtime validation of pointers. Tags are 4 bits in size applied with a granularity of 16 bytes. The maximum heap size is configured to $2^{34}$ bytes (16GB).}, label=lst:asmpointervalidation, float=t, belowskip=-1.2 \baselineskip]
rorx  rcx, ptr, 34          // extract tag to cl
shrx  rax, rcx, shadow_base // map ptr to shadow
cmp   cl,  [shadow_base + rax]
jnz   .fallback_check       // might be non-heap
fallback_succ:
...
fallback_check:
xor   ptr, heap_start
shr   ptr, 38         // iff heap: result is 0
jnz   .fallback_succ  // jumps if non-heap ptr
ud2                   // crash
\end{lstlisting}

\paragraph{Handling the Fork System Call}%
\label{sec:forksupport}
Our prototype implementation relies on using \emph{mmap} with the \texttt{MAP\_SHARED} flag to map the same physical page to different virtual addresses.
The current Linux kernel does not support mapping process-private pages multiple times into the address space.
As a consequence, our allocator changes the semantics of the \emph{fork} system call.
Since shared pages are also shared between parent and child after a fork, all heap-allocated data is shared between parent and child in our current prototype.
This is a limitation of the memory management APIs of the Linux kernel.
However, prior work has shown that this can be handled in userspace by instrumenting the \emph{fork} system call~\cite{oscar2017}.
Here, the process manually creates copies of the shared pages while forking such that parent and child have different physical pages backing the heap allocations.
Alternatively, the kernel could be modified to provide an API supporting aliased non-shared page mappings efficiently.
Both approaches are straightforward to adopt in our prototype.
For the evaluation, we ensured that the benchmarked applications do not utilize the \emph{fork} system call.

\paragraph{Non-standard Pointer Usage}
In general, our metadata embedding scheme is highly compatible with
\begin{inparaenum}[1.]
	\item system calls,
	\item non-instrumented libraries, and
	\item inline assembler code.
\end{inparaenum}
Hence, the \uaf mitigation scheme is by design compatible with legacy code.
However, the optimization techniques described in \Cref{sec:optimizing} potentially lead to false alarms in case pointers are handled in an uncommon fashion.
For example, passing a freed pointer across function boundaries does not necessarily constitute a \uaf considering that the callee might never dereference the pointer.
However, this behavior is incompatible with our \emph{validating in the caller} optimization.
Furthermore, indexing an out-of-bound pointer such that the accessed address is correct regarding allocation boundaries is as well not compatible with our \emph{partially-aliasing pointer} optimization.
We noticed a very few cases where this issue becomes apparent, namely the \emph{perlbench} SPEC benchmark, for which we had to disable our optimizations.
By doing so, we avoid false alarms at the cost of performance overhead increase but still ensuring the same security guarantees.
For two other benchmarks, \emph{gcc} and \emph{xalancbmk}, we noticed that a few functions (four functions and three function templates respectively) are not compatible to our \emph{validating in the caller} optimization.
Instead of disabling our optimization entirely for these programs, we simply move pointer verification from callers to callees for the few non-compliant functions.
For all other benchmarks, we did not observe any unusual pointer usages.
For a full list of incompatible functions in SPEC 2017 intspeed and a discussion of omitted instrumentation in Firefox we refer to \Cref{appendix:ignorelist}.

\section{Performance Evaluation}%
\label{sec:evaluation}

\paragraph{Setup}
Our evaluation on the SPEC CPU2017 intspeed benchmark runs on server hardware, equipped with an Intel Xeon Gold 6230R CPU and \SI{196}{\giga\byte} DDR4 RAM, running Ubuntu 20.10.
Both hyper-threading and Intel Turbo Boost are disabled for increased reproducibility.
We also evaluate our mitigation against multiple browser benchmarks.
This evaluation runs on system with an Intel i7-10510U CPU, \SI{32}{\giga\byte} of DDR4 RAM, and running Ubuntu 20.10.
In contrast to the evaluation on SPEC CPU2017, hyper-threading and Intel Turbo Boost are enabled as this is the most realistic scenario for browser execution.
All SPEC benchmarks leverage profile-guided optimization and ThinLTO~\cite{clanglto} for peak performance.
We omit the \emph{exchange} benchmark from evaluation as we do not support instrumentation of Fortran code.
We utilize the \emph{Clang} compiler version~\version{10}.
The allocator for reference values of SPEC CPU2017 intspeed is \emph{mimalloc} \version{1.6.1} with transparent huge pages (THP) enabled.
The bars for evaluating SPEC show the median overhead in percent relative to the reference execution running with the \emph{mimalloc}.
Error bars indicate minimum and maximum overhead of three runs relative to the same baseline.

Benchmarks on Firefox~81 utilize profile-guided optimization and ThinLTO.
For all Firefox benchmarks, we configure \texttt{--enable-valgrind} and \texttt{--disable-sandbox} (for compatibility with MarkUs) as well as \texttt{--disable-forkserver} (for compatibility with our prototype, see \Cref{sec:forksupport}).
Note that valgrind is not actually running, however the flag is required for build compatibility with MarkUs.
Evaluating Firefox with MarkUs~\cite{markus2020}, ASAN~\cite{serebryany2012asan}, and our prototype requires \texttt{--disable-jemalloc}, as all of them utilize a modified allocator.
As a result, our allocator handles all allocations but the garbage-collected objects of the JavaScript engine.
The baseline build does not use this flag and runs with the built-in \emph{jemalloc} allocator.

We compare \toolname\ to two other \uaf\ mitigations: MarkUs~\cite{markus2020}, the fastest and most recently published \uaf\ mitigation scheme, and the sanitizer ASAN.
We cannot compare \toolname\ with other \uaf\ mitigations as most prototypes, like DangSan~\cite{dangsan2017} and CRCount~\cite{crcount2019}, are built upon very old LLVM versions (i.e., version 3.8).
As a result, we cannot compare on the SPEC benchmark because simply changing to a new LLVM version and a newer allocator increases performance and biases the results towards our solution, which is based on LLVM 10. %

\paragraph{Runtime Overhead}
We evaluate the runtime overhead %
on the SPEC CPU 2017 intspeed benchmarks, the results are summarized in \Cref{fig:RTOverhead}.
We omitted ASAN because the geomean overhead is more than \SI{123}{\percent} (see \Cref{appendix:fullrtoverhead} for an overview of ASAN's overhead).

\begin{figure}[t]
	\centering
	\includegraphics{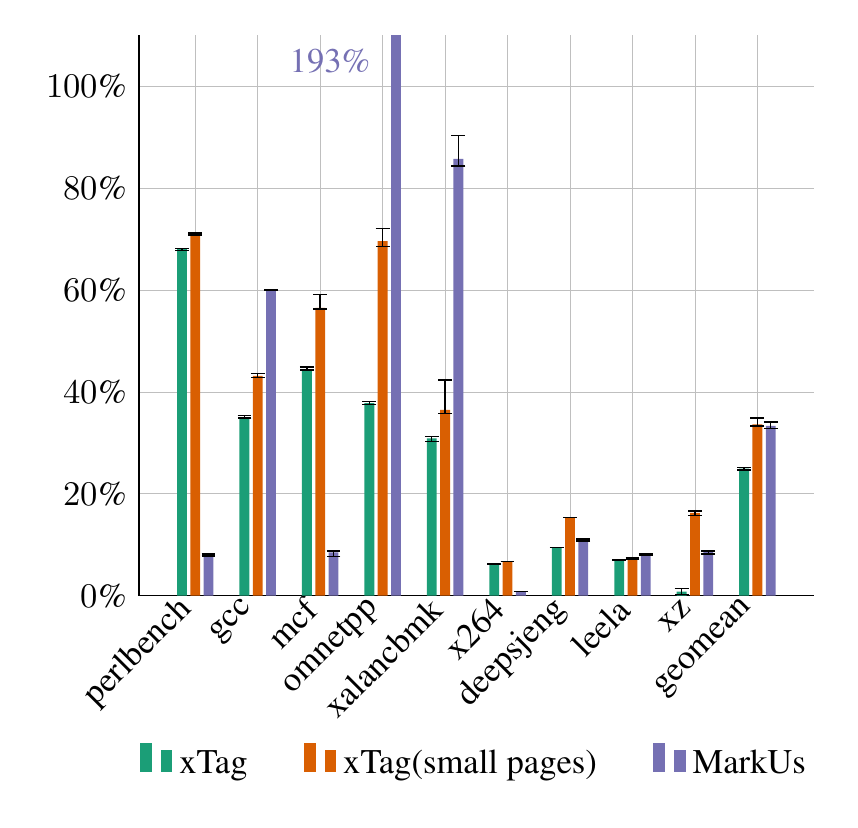}
	\caption{Runtime overhead relative to a baseline with \emph{mimalloc} \version{v1.6.1} (THP enabled) on SPEC CPU2017 intspeed.}%
	\label{fig:RTOverhead}
\end{figure}

While our measurements of MarkUs on the individual benchmarks are generally aligned with the values reported by the authors in the paper~\cite{markus2020}, we noticed two differences.
First, the \emph{gcc} benchmark crashes, we show the overhead from the original publication~\cite{markus2020} instead.
Second, the reported overhead on \emph{omnetpp} is reported as roughly \SI{110}{\percent}, while we measure \SI{193.3}{\percent} overhead on our system.

\begin{figure*}[t]
	\begin{subfigure}{.33\linewidth}
		\centering
		\includegraphics[width=\textwidth]{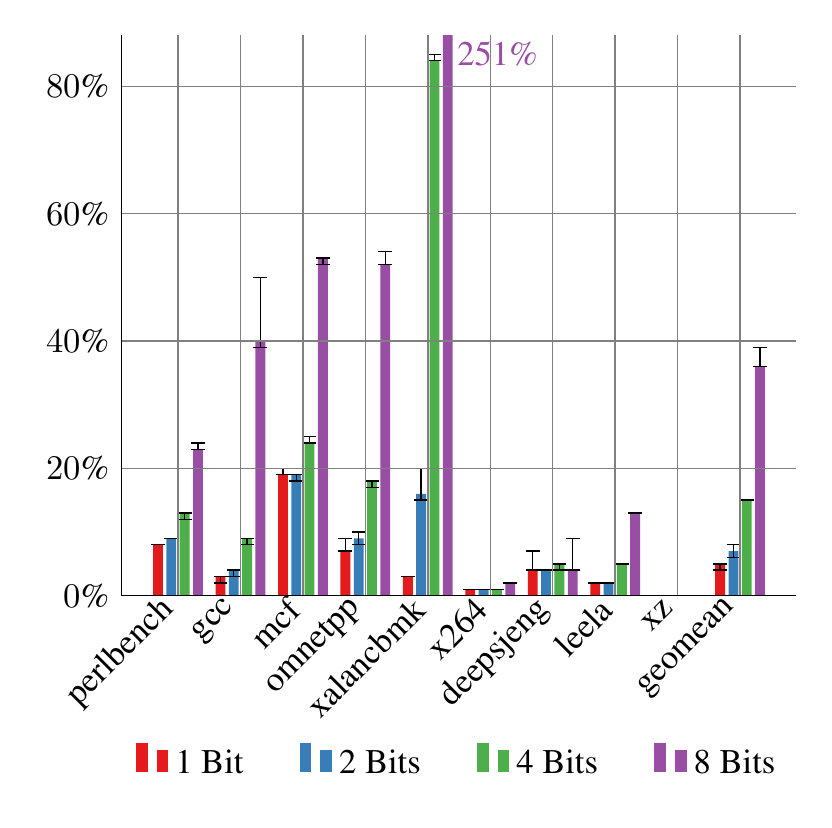}
		\caption{Tags are selected randomly.}%
		\label{fig:allocoverhead:rand_huge}
	\end{subfigure}
	\begin{subfigure}{.33\linewidth}
		\centering
		\includegraphics[width=\textwidth]{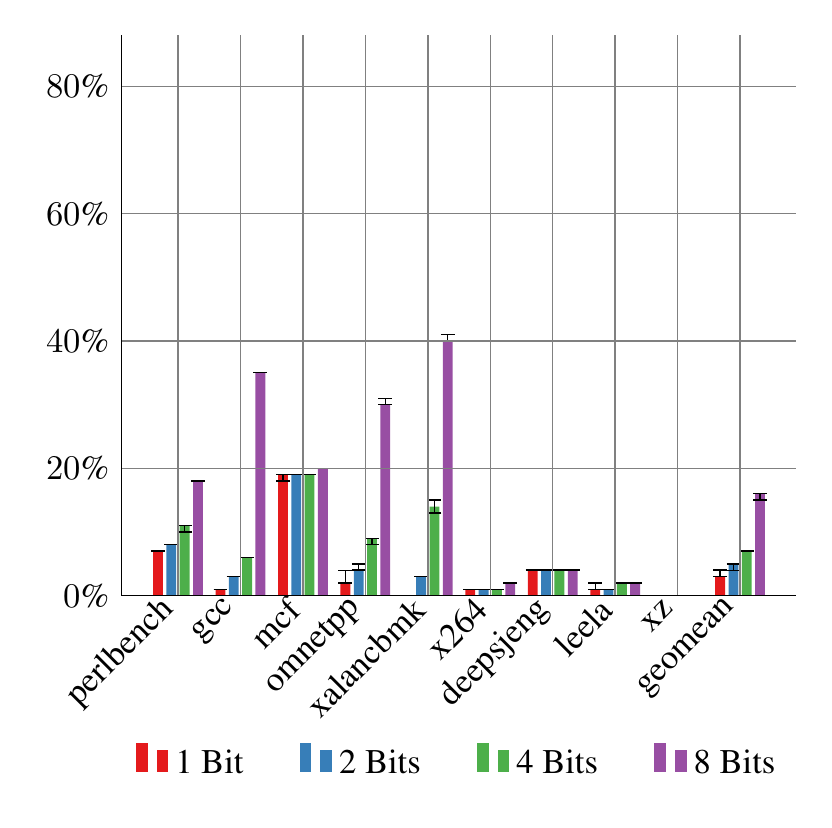}
		\caption{Tags are selected in a generational fashion.}%
		\label{fig:allocoverhead:gen_huge}
	\end{subfigure}
	\begin{subfigure}{.33\linewidth}
		\centering
		\includegraphics[width=\textwidth]{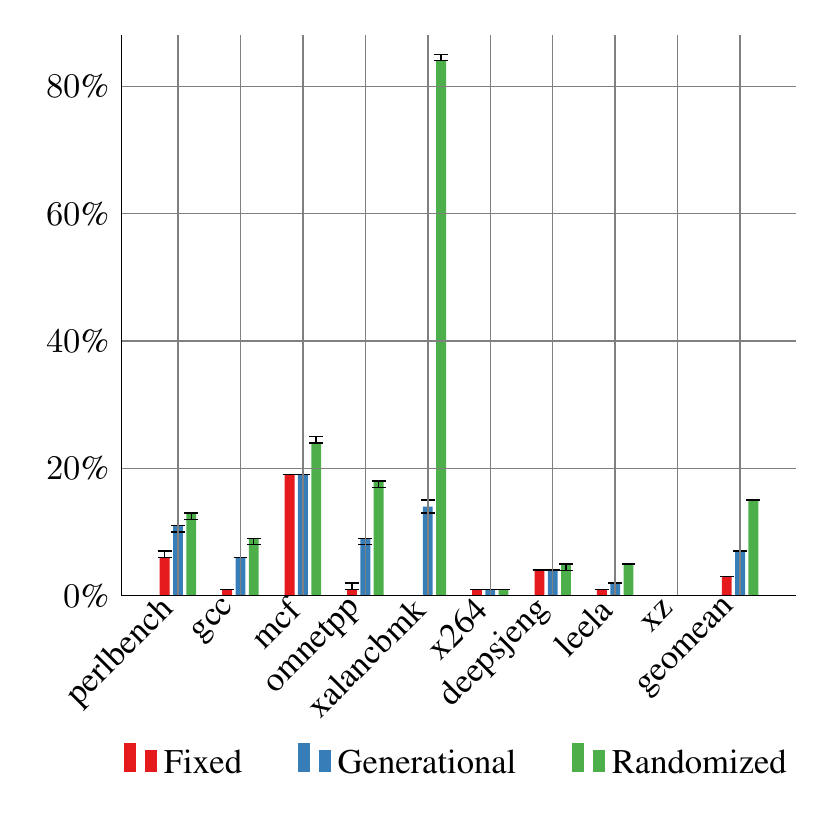}
		\caption{Comparison of tagging strategies with a tag size of 4 bit.}%
		\label{fig:allocoverhead:comparison_huge}
	\end{subfigure}

	\caption{\allocname induced performance overhead with \SI{2}{\mega\byte} pages on SPEC CPU2017. Different tag sizes and tagging strategies are compared to  \emph{mimalloc} \version{v1.6.1} (THP enabled).}
	\label{fig:allocoverhead}
\end{figure*}

In most benchmarks, \toolname\ induces a smaller runtime overhead than MarkUs, showing that \toolname\ performs better compared to the current state-of-the-art \uaf\ mitigation.
The geomean overhead for \toolname is \runtimeOverhead, while it is \SI{33.3}{\percent} for MarkUs, a \relativeImprovementMarkUs relative reduction.
Reducing the TLB overhead incurred by \allocname with huge pages is evidently critical for the performance of \toolname and further evaluated in \Cref{fig:allocoverhead}.
Moreover, the benchmark with SPEC CPU runs on a single CPU core, which biases the results towards MarkUs:
Note that MarkUs utilizes a separate thread for parts of the garbage collection process and as such can potentially utilize a second CPU core to its advantage.
For high-throughput workloads, where all cores run at full capacity, the overhead induced by the parallel garbage collection might increase further.
In contrast, \toolname's overhead is fully deterministic and restricted to a single process.
We show the measurement results for the overhead on SPEC CPU intspeed \emph{without} the parallel garbage collection in \Cref{appendix:serialperformance}, the geomean overhead for MarkUs is then even \SI{39.7}{\percent}.

\begin{figure}[h]
	\centering
	\includegraphics{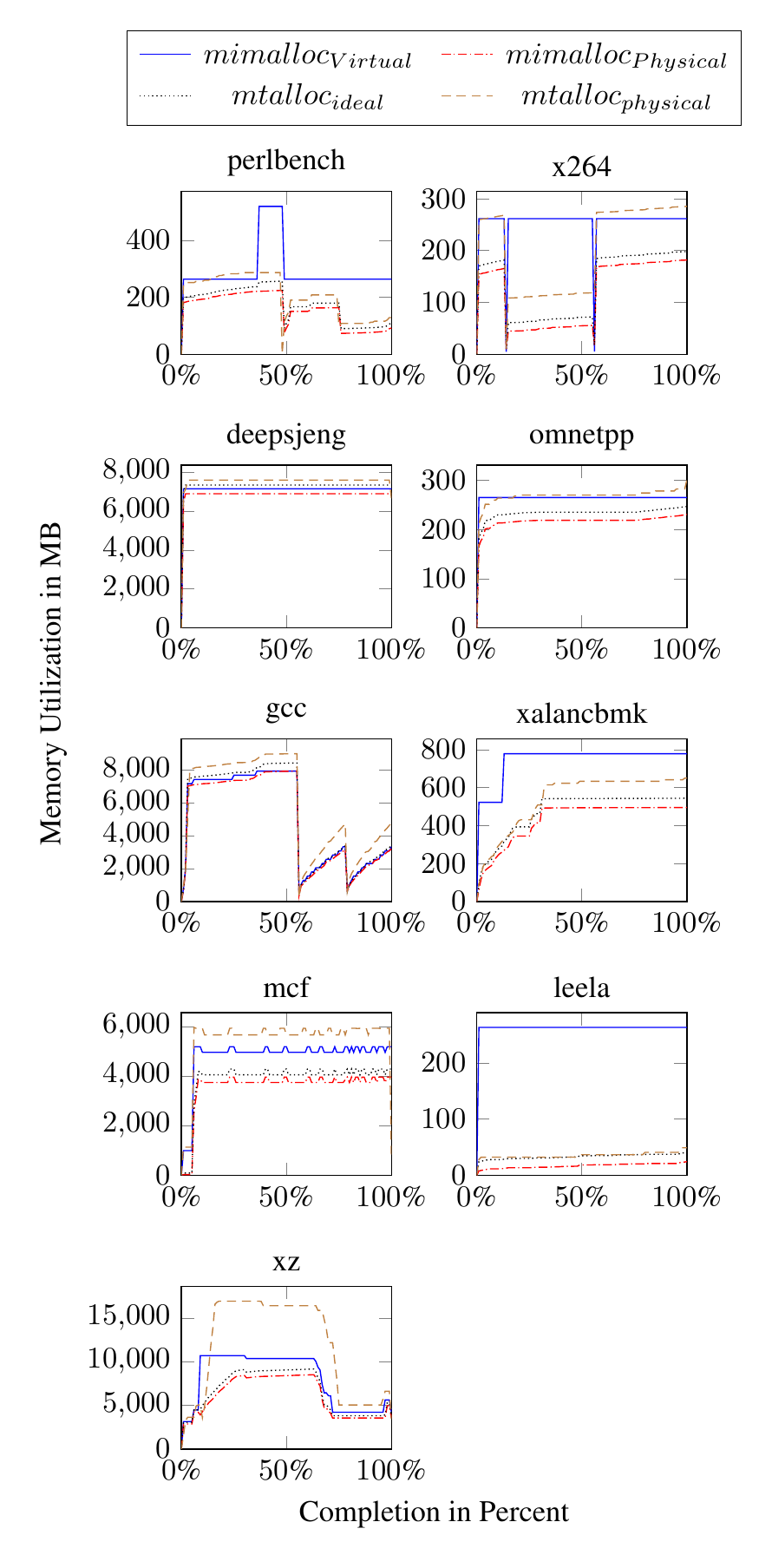}
	\caption{Memory consumption of \allocname with \SI{2}{\mega\byte} pages for SPEC CPU2017 intspeed benchmarks.}
	\label{fig:appendix:memovertime}
\end{figure}

\paragraph{Memory Overhead}
As our tagging scheme requires 1 tag byte per \SI{16}{\byte} of allocated virtual memory, we expect about \SI{6.25}{\percent} memory overhead.
Assume a program allocates \SI{1}{\mega\byte} heap memory with \allocname.
The physical memory backing the allocation is lazily assigned by the OS, i.e., assignment is deferred until the allocated memory gets accessed.
However, tags are stored in shadow memory at allocation time.
Hence, we immediately require $1 \si{\mega\byte} / 16 = 64 \si{\kilo\byte}$ of physical memory for tag storage.
\Cref{fig:appendix:memovertime} shows the memory overhead for the SPEC CPU2017 benchmark, which we omitted in \Cref{sec:evaluation} due to space constraints.
Virtual memory is sampled from \texttt{/proc/pid/stat}, whereas physical memory usage is profiled by placing the processes in a dedicated \texttt{cgroup} and sampled from \texttt{memory.usage\_in\_bytes}.
The baseline execution uses an unmodified \emph{mimalloc} \version{v1.6.1} as allocator.
Assuming memory overhead stems from tags in shadow memory only, we compute idealized memory usage of \allocname with $\allocname_{Ideal} = mimalloc_{Physical} + mimalloc_{Virtual} / 16$\\
Physical memory usage of \allocname with \SI{2}{\mega\byte} pages assigning 4 bit tags with generational tag selection is profiled in $\allocname_{physical}$.
Decreasing the page size to \SI{4}{\kilo\byte} slightly decreases physical memory utilization.
While memory overhead generally corresponds to the expected behavior for 6 out of 9 benchmarks, \emph{xz} and \emph{mcf} show an increase of \SI{60}\percent.
\emph{x264} exhibits an initial overhead of \SI{60}\percent, execution terminates with a memory overhead of \SI{70}\percent.
Ideally, physical memory overhead of \allocname arises due to tag storage only.
However, we also identified increased kernel memory for managing virtual memory mappings as well as aggressive page retainment by \allocname for reducing system calls as additional sources for overhead.
We believe that the latter could be optimized by introducing kernel support.

\paragraph{Measuring TLB Pressure}
As described in \Cref{sec:tlboptimization}, embedding a tag diversifies the virtual addresses and hence increases pressure on the TLB.
To quantify the performance penalty, we evaluated different tag sizes on the SPEC benchmark, the results can be found in \Cref{fig:allocoverhead}.
We measured the performance overhead for tags of size 1, 2, 4, and 8 \si{\bit} using different tagging strategies and a page size of \SI{2}{\mega\byte}.
A similar evaluation of \SI{4}{\kilo\byte} pages is deferred to \Cref{appendix:4kbpages}.
Remember that the two hardware-backed memory tagging solutions ARM MTE and SPARC ADI use 4 tag bits.
Note that the penalty includes some overhead due to the increased number of system calls.
To measure the system call overhead, we introduced a \emph{fixed} tagging strategy that always uses the same tag value and thus does not increase TLB pressure.
The baseline of our measurement is a profile-optimized build running with an unmodified version of \emph{mimalloc}.
We cannot use the standard \emph{glib} allocator as baseline since this could hide some overhead introduced by \allocname.
The individual benchmarks are not instrumented for any kind of runtime mitigation as we are interested in the allocator overhead only.

Our measurements show that using \SI{8}{\bit} tags is indeed impractical, the TLB pressure and system call overhead tremendously increases.
As is shown in \Cref{fig:allocoverhead:rand_huge}, a random tagging strategy with \SI{8}{\bit} tags induces up to \SI{251}{\percent} overhead in the worst-case benchmark.
On the same benchmark, a \SI{4}{\bit} random tag induces only \SI{84}{\percent} overhead.
Even though our generational tagging strategy can mitigate some TLB pressure, we conclude that \SI{4}{\bit} tags are currently the only tag size that offers practical performance implications, while still providing acceptable security guarantees.

We also measure the impact of our generational tagging strategy on the induced TLB pressure.
For example, the worst-case benchmark (xalanc) with \SI{8}{\bit} drops from \SI{251}{\percent} with random tag selection to \SI{40}{\percent} with generational tag selection.
\Cref{fig:allocoverhead:comparison_huge} shows a comparison between the generational and randomized tagging strategies.
Overall, our generational tagging strategy can reduce the runtime overhead by reducing TLB pressure to acceptable numbers, e.g., the xalanc benchmark with 4 bit tags can be reduced from \SI{84}{\percent} to \SI{14}{\percent}.

Lastly, we measure system call overhead and TLB pressure with the results shown in \Cref{fig:allocoverhead:comparison_huge}.
For this, we compare three tagging strategies: randomized, generational and fixed.
The fixed tagging strategy allows us to measure system call-induced overhead.
Since we always use the same tag, there is no impact on the TLB.
As such, we measure only the overhead induced by our changes to the baseline allocator, such as additional system calls.
We can see that in most benchmarks this part of the overhead is negligible but on certain benchmarks (omnetpp and xalanc), with a lot of heap allocations, it induces a performance impact. The majority of the overhead is indeed caused by the TLB pressure.

\begin{figure}[t]
	\centering
	\includegraphics[width=\columnwidth]{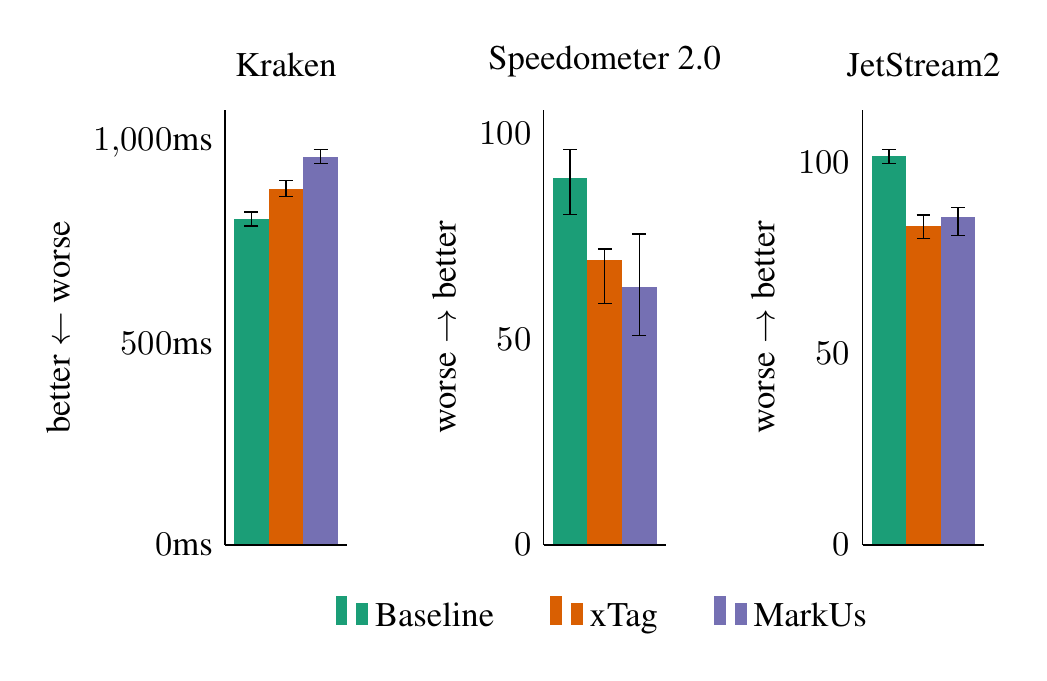}
	\caption{Performance overhead for three browser benchmarks. Note that Kraken measures performance in ms (lower is better), while the other two show performance impact (higher is better).}%
	\label{fig:browserperf}
\end{figure}

\paragraph{Browser Performance Overhead}
To evaluate the performance of our mitigation on browser workloads, we run three popular web benchmarks: \emph{Kraken}, \emph{Speedometer 2.0}, and \emph{JetStream2}.
\Cref{fig:browserperf} shows the benchmark results of the Firefox web browser protected with \toolname, MarkUs, and an unprotected build as baseline.
The \emph{Kraken} benchmark measures performance of JavaScript execution.
As the results are in $ms$, lower results are better.
Repeating the measurement 10 times for increased accuracy reveals a penalty of 9\% for \toolname compared to 18\% for MarkUs.
The \emph{Speedometer 2.0} benchmark was run three times, while taking into account all 10 repetitions reported per run.
This benchmark simulates user interaction and aims to quantify responsiveness.
The median penalty of \toolname is 29\%, compared to 42\% from MarkUs.
The last benchmark, \emph{JetStream2}, is a browser benchmark including JavaScript and WebAssembly, again executed 10 times.
Our performance penalty of 22\% for \toolname is just slightly more than 18\% for MarkUs.

\section{Exploit Case Studies}%
\label{sec:casestudy}

To show that we can effectively detect \uaf\ conditions, we use three recent real-world vulnerabilities in the Firefox web browser to demonstrate that our approach can handle complex, real-world applications.
All three vulnerabilities are publicly documented and come with a test case to reproduce the \uaf vulnerability.
We built an instrumented version of the Firefox browser in version 65 and ran the test cases.
For all three case studies, our instrumentation can successfully detect and prevent an attempt to exploit the \uaf vulnerability.

\paragraph{CVE-2018-18500} is a UAF vulnerability in the HTML5 parser of Firefox.
Custom HTML elements might be freed during parsing, while still being in use.
The destructor of \texttt{nsHtml5StreamParser} frees an \texttt{nsHtml5TreeBuilder} object,
even though it is subsequently used in the function \texttt{nsHtml5TreeOperation::Perform}.
Our mitigation terminates program execution before accessing the freed memory, hence successfully preventing an exploitation of the bug.

\paragraph{CVE-2019-11752} is a UAF in the IndexedDB implementation of Firefox.
A key value deleted in \texttt{IDBObjectStore::DeleteIndex} might be accessed later on during a call to the function \texttt{indexedDB:: KeyPath::ExtractKey}.
Our instrumentation successfully terminates program execution in the latter function, hence mitigating the vulnerability.

\paragraph{CVE-2019-11691} is a UAF vulnerability occurring during processing of \texttt{XMLHttpRequests} in combination with event loops.
The garbage collector reclaims objects not kept alive while still being used.
This causes an access to freed memory in
\texttt{XMLHttpRequestMainThread::DispatchProgress Event}.
Our mitigation is again able to detect the access and terminates program execution accordingly.

\section{Security Considerations}%
\label{sec:securityanalysis}
In the following, we discuss the security implications of \toolname\ to ASLR, the effectiveness of a 4-bit tag, as well as multithreading aspects.

\paragraph{ASLR Weakening}
From a theoretical point of view, our pointer tagging scheme potentially weakens address space layout randomization (ASLR).
Given that we have multiple mappings of the same data in the virtual address space, our method reduces the potential entropy by \SI{4}{\bit}.
The pointer prefix allowing to quickly distinguish between tagged and untagged pointers changes with every execution due to randomization by the operating system.
The pointer format shown in \Cref{fig:heappointer} features \SI{10}{\bit} of randomness in $Prefix$ due to randomization by the OS. Furthermore, the allocator can randomize the $Suffix$ bits, allowing for up to additional 34 bits of randomness, modulo alignment requirements of individual allocations.

However, we argue that in practice, our pointer tagging scheme does not have any significant impact on ASLR.
Current ASLR implementations on all major unhardened operating systems do not actually utilize the full address space for randomization.
For example, a typical unhardened Linux system utilizes the same pointer prefix for all \emph{mmap} allocated memory and a different fixed prefix for all \emph{brk} allocated data.
This prefix is constant anyway, so our embedded metadata does not actually reduce pointer entropy when compared to a stock Linux kernel.
Furthermore, brute-force attacks against ASLR~\cite{Shacham2004aslr} have become less of an issue in recent years.
Already on \SI{32}{\bit} systems, brute-force attacks required a long time to succeed.
As such, arbitrary~\cite{snow2013jitrop,rogowski2017browserdop} or limited~\cite{morton2018webserverdop} read exploit primitives have become an integral part of many real-world exploit chains to bypass ASLR reliably.
Additionally, recent micro-architectural and timing side-channel attacks~\cite{oren2015jscache,gras2017anc,canella2020kaslr} have shown that ASLR does not offer a strong defense against local attackers with limited code execution (i.e., in the browser threat model).
As such, we believe the potential entropy reduction in ASLR induced by our pointer tagging scheme does not induce any practically relevant disadvantage.

\paragraph{Tag Reuse}
As other hardware-backed memory tagging solutions, our implementation utilizes \SI{4}{\bit} tags.
This allows for an efficient implementation and low runtime overhead.
However, as this results in only 16 different tags, there is a small probability of a \emph{tag reuse} condition.
Assuming there exists a dangling pointer with a specific tag, after multiple allocations and free operations, the memory the dangling pointer refers to could potentially be re-allocated with the same tag again.
Then the \uaf\ condition would remain undetected.
Note that due to random tag selection, an adversary cannot exploit deterministic behavior~\cite{sotirov2007heap} to deliberately create colliding tags.
With a $1/16$ chance of tag re-use, we believe this to be improbable enough in practice that \uaf\ will not be missed.
Furthermore, other memory tagging schemes, such as ARM MTE and SPARC ADI, also utilize the same tag size and as such are also prone to tag reuse.

However, more problematic in an adversarial setting is a potential brute-force attack against the tag value.
Under normal conditions, the process terminates whenever a \uaf\ is detected.
As such, there is a $1/16$ chance that the \uaf\ exploit succeeds and the next attempt would utilize new random tag values.
However, if the attacker is able to perform the \uaf\ access in a crash-resistant manner~\cite{gawlik2016crashresistance}, then it is possible to brute-force the tag.
The attacker needs to be able to force the target program to re-allocate the same memory under a new tag.
The attacker probes the tag by triggering access through the dangling pointer.
If a crash is observed, the attacker again forces a re-allocation under a new tag and probes again.
An adversary with access to a crash-resistance primitive can bypass any existing tagging-based solution.
In such a scenario, even 8-bit tags do not offer significant advantage over 4-bit tags.
That is, this is not a limitation of our implementation, but an open problem of any tagging-based scheme.

Prior work has shown that there are multiple ways to achieve crash-resistant exploit primitives~\cite{gawlik2016crashresistance,kollenda2017autocrop}.
However, most of the known crash-resistance primitives are not suited to brute-forcing our pointer tagging scheme.
For example, primitives that abuse the kernel as a confused deputy to dereference a pointer cannot be used to gain information about the actual tag value, as every possible embedded tag can always be dereferenced by the kernel.
Only crash-resistance primitives that mask all signals (i.e., also trapping instructions) can be utilized for brute-forcing the memory tag and those are rare to find~\cite{kollenda2017autocrop}.
Crash-resistance is a big problem to any exploit mitigation as it enables probing the address space, e.g., to find hidden shadow stacks in CFI schemes.
As such, \toolname---similar to any other mitigation---will benefit from software systems removing crash-resistance primitives.

\paragraph{Multithreading}
\label{sec:securityanalysis:multithreading}
In general, our solution supports multithreading, which is challenging to support securely in exploit mitigations~\cite{Conti2015losingcontrol,bialekoffensivecon}.
Similar to other non-atomic software-only mitigations (e.g., Microsoft's Return Flow Guard~\cite{bialekoffensivecon}), there is a potential time-of-check to time-of-use race condition inside the \uaf\ check.
A memory object is freed by one thread, while at the same time another thread accesses the same memory object.
Now in the instrumentation, there is a small race window between the shadow memory load and the actual memory load.
This means while one thread is freeing a pointer, a second thread could still access the same memory with the previous pointer tag (since the shadow memory has not been updated yet).
While this race window is generally very small, it could be lengthened by, e.g., compiler optimization that move the instrumentation code away from the actual pointer dereference.
This race window exists in any software-only mitigation.
For detecting \uaf\ we believe that 
\begin{inparaenum}[(1)]
	\item this race window is small enough to be practically irrelevant, and
	\item concurrent \uaf\ are sufficiently rare, as usually concurrent access is protected by locking mechanisms in the application. 
\end{inparaenum}
In our exploit case studies (see \Cref{sec:casestudy}) we did not encounter any concurrent \uaf\ conditions.

\section{Related Work}
\label{sec:relatedWork}

In the past years, memory safety attacks and defenses were a very active research field.
We now discuss how our approach relates to prior work in this area.

\paragraph{Automatic Memory Management}
In principle, automatic memory management basically eradicates \uaf\ exploits by providing temporal memory safety.
Although garbage collectors for the C and \Cpp languages were proposed and implemented~\cite{Boehm1991gc}, none of them has found widespread use in practice.
In general, garbage collection in C/\Cpp is not considered safe due to potential pointer hiding issues~\cite{Boehm1996gcsafety} in the application.

Recently, MarkUs~\cite{markus2020}, a hybrid approach of manual memory management and garbage collection, has been proposed as a security defense.
MarkUs is implemented on top of the Boehm garbage collector, which features a parallel marking algorithm.
The basic idea is to apply garbage collection to all objects that have been marked as freeable by a call to the standard \emph{free} function.
However, as long as a dangling pointer is detected by the garbage collector in memory, the object is not freed.
This hybrid approach is semantically secure, as no undefined behavior is introduced by freeing objects, whose pointer references are hidden from the garbage collector.
Only those objects that are also manually freed are collected by the garbage collector.
However, this also means that this approach cannot identify complex \uaf\ conditions where the dangling pointer is a hidden pointer.
In contrast, a pointer tagging based solution such as ours can detect \uaf\ conditions even for hidden pointers:
As long as the pointer-embedded tag is preserved, the pointer hiding requires no changes.
Furthermore, we show experimentally that \toolname\ offers better performance characteristics than MarkUs in many workloads (see \Cref{sec:evaluation}) and we believe that \toolname\ is better suited for high-throughput scenarios.

Reference counting is another approach to automatic memory management, which is becoming more popular in modern \Cpp codebases with smart pointers.
Smart pointers encapsulate a raw pointer together with a reference counter.
However, smart pointers must be manually inserted into the source code.
CRCount is an automatic solution that can be used to retrofit reference counting into legacy C/\Cpp codebases~\cite{crcount2019}.
However, solutions that automatically introduce reference counting are prone to incorrect frees and potential errors when non-instrumented code, like inline assembly, copies the pointer without increasing the reference counter.
In contrast, \toolname\ handles non-instrumented code gracefully.

\paragraph{Pointer Invalidation}
Several works have studied a different approach to \uaf\ mitigation.
Instead of changing some property of the allocation, such as the shadow memory tag in this paper, dangling pointers can be explicitly invalidated~\cite{Lee2015dangnull,freesentry2015,dangsan2017}.
This solves the root cause of \uaf, i.e., dangling pointers are eradicated.
However, to achieve this property, all pointers must be closely tracked so that the instrumentation can determine which pointers need to be invalidated.
This leads to expensive instrumentation and metadata tracking.
A similar approach is taken by the Oscar allocator as well as by FFMalloc ~\cite{oscar2017, 2021ffmalloc}.
However, instead of tracking pointers, the allocator never reuses virtual addresses.
After a free, the dangling pointer points to a memory area never assigned to another allocation, hence preventing any attempt to corrupt allocated data.
This allows to perform \uaf\ checks without instrumenting the code.
Similar to \toolname, Oscar and FFMalloc repurposes the MMU for security purposes.
However, in contrast to \toolname\, such techniques use a lot of virtual addresses.
As a consequence, this increases TLB pressure, but also leads to an increased system call overhead.

\paragraph{Preserving Type Safety}
Another approach to limiting the impact of \uaf\ bugs is the use of type-pooled allocators.
By allocating only the same types of objects within one heap region, we can preserve type safety even under \uaf\ conditions~\cite{dhurjati2003memorysafetywithoutruntimechecks,Kouwe2018type,Periklis2010Cling}.
This prevents an attacker from breaking type safety as part of the exploit.
For example, a common strategy during heap exploitation is to force the application to interpret attacker-controlled input as a pointer value.
By preserving type safety, such exploit strategies would not be possible anymore.
However, it is not clear whether such type-pooled allocators can be efficiently and securely implemented in practice, especially for complex \Cpp programs relying on custom allocators and in-place construction of objects via placement new.
Several web browsers have implemented some form of pool-based allocation, most notably Chrome's \emph{PartitionAlloc}~\cite{chromepartitionalloc}.
However, a recent analysis of serious security bugs in Chrome~\cite{chromememsafety} revealed that about \SI{36}{\percent} of the analyzed 912 high or critical severity security bugs since 2015 were related to \uaf.
This indicates that even in the face of hardened allocators, \uaf\ is still a major problem.

\paragraph{Delayed Free Lists}
To make \uaf\ conditions harder to exploit, many hardened allocators~\cite{Berger2006diehard,Novark2010dieharder,Sam2017freeguard,2018guarder}, such as scudo~\cite{scudo}, or ASAN's allocator use a delayed free list.
This means that any heap allocated object is not immediately freed, but is stored in a queue instead.
At some point during the program's execution, the allocator decides to free the objects within the delayed free list.
This helps to prevent exploitation of \uaf\ conditions by introducing a non-determinism in the allocation order.
However, such delayed free lists are not as useful under memory pressure, which might be attacker-controlled.
For example, if the attacker has a lot of control of the allocations performed by the target, e.g., in the web browser setting, the attacker is likely able to fill the delayed free list up to a point where the desired object is freed, hence forcing a free.
As such, even probabilistic defenses built around pointer tagging offer better security properties than introducing non-deterministic allocation behavior.

\section{Conclusion and Future Work}

In this paper, we introduced a practical pointer tagging scheme, called \toolname, for the Intel x86-64 architecture that overcomes the lacking hardware support on this platform. By embedding the tag bits in a clever way, we obtain a scheme that is highly compatible and performant, we consider this scheme to be the first efficient and effective pointer tagging scheme for x86-64. Based on \toolname and several optimizations that significantly reduce TLB pressure, remove redundant checks, and reduce memory overhead,
we showed how a \uaf and a double-free mitigation scheme can be implemented on top of this method. Compared to state-of-the-art methods, our approach leads to a lower performance impact, while offering strong security properties.

The current implementation of \toolname lays the foundation for software-based memory tagging.
Note that out-of-bound memory accesses are an orthogonal class of memory safety violations.
Our proposed scheme can be extended to cover out-of-bounds memory violations as follows:
First, this requires to replace generational tag selection with randomized tag selection, as adjacent allocations must feature (probabilistically) different tags.
The performance impact of this modification due to increased TLB pressure is quantified in~\Cref{fig:allocoverhead:comparison_huge}.
Second, the partially-aliasing pointer optimization is no longer applicable.
Instead, a naive implementation to protect against out-of-bounds violations could validate all accesses to structs and array separately.
A more advanced implementation could reduce the performance overhead by reasoning over allocation bounds.
As an example, iterating over an array with known size could be preceded by a check verifying that the first and last element are within bounds.
Assuming the check is successful, all inner elements must also be within the bounds.
This is in stark contrast to hardened allocators, which instead seek to reduce the consequences of memory corruptions after the fact, rather than preventing them in the first place.

\section*{Acknowledgments}
Funded by the Deutsche Forschungsgemeinschaft (DFG, German Research Foundation) under Germany's Excellence Strategy - EXC 2092 CASA - 390781972.

\bibliographystyle{IEEEtranS}
\bibliography{references}

\appendices

\section{Optimization Ablation Study}
\begin{figure}[t]
    \centering
    \includegraphics{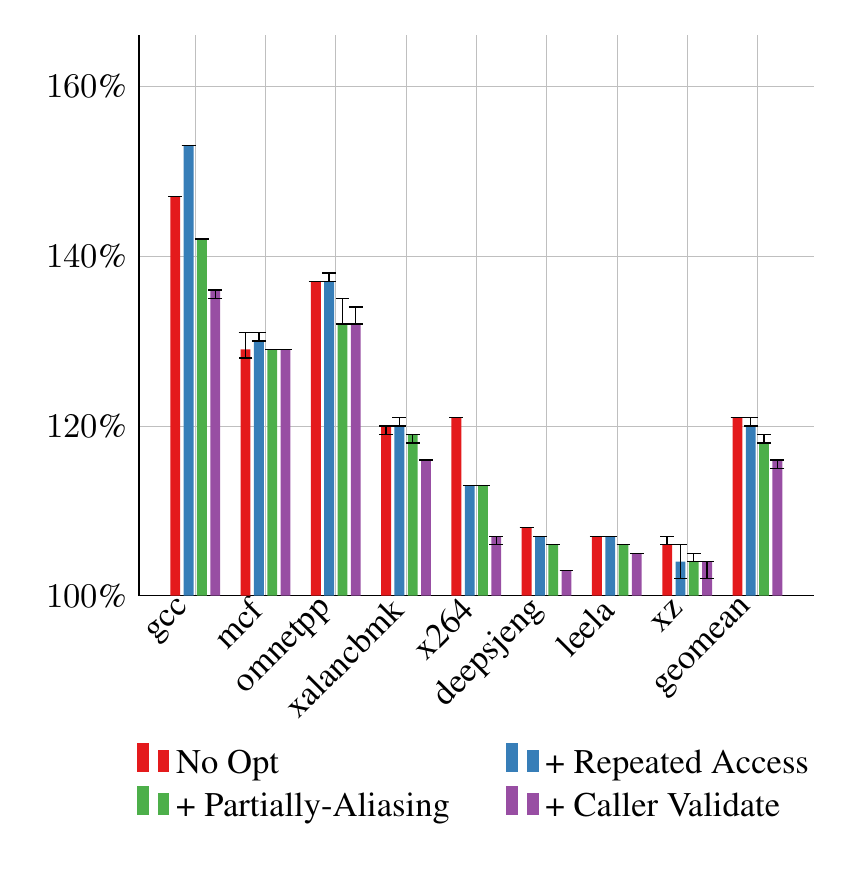}
    \caption{Performance overhead of \toolname relative to a baseline disabling all runtime verification. The allocator configuration is generational tagging, \SI{4}{\bit} tags, and THP enabled.}%
    \label{fig:Ablation}
\end{figure}

To better understand the individual compiler optimizations described in \Cref{sec:optimizing}, we conducted an ablation study.
We executed the SPEC CPU2017 benchmark with generational tag selection and huge pages enabled, while gradualy enabling the \toolname-specific compiler optimizations.
The performance overhead is measured relative to a baseline execution using \allocname with the same configuration (generational tagging, \SI{4}{\bit} tags, and THP enabled), but with all runtime checks disabled.
As a result, we only measure the overhead of the runtime check inserted by \toolname.
Note that \emph{perlbench} has been removed from the benchmark as it is incompatible with some of the tested optimizations (see \Cref{sec:implementation}).
The results of the study are shown in \Cref{fig:Ablation}.
The study shows that all optimizations contribute a significant reduction in overall geomean performance overhead.
In total, the overhead introduced by runtime checking decreases from 21.1\% (with no optimizations) to 15.8\% (all optimizations enabled).
There are two benchmarks (\emph{gcc} and \emph{mcf}) which suffer from a performance degradation when comparing the baseline execution with the \emph{repeated access via same pointer} optimization.
We explain this difference by the optimization strategy that inserts checks in places with worse runtime properties.
Since the repeated access optimization is a prerequisite for the partially-aliasing optimization, we conclude that it is still worth using in any case.

\section{\allocname Overhead on small pages}
\label{appendix:4kbpages}

\begin{figure*}[t]
	\begin{subfigure}{.33\linewidth}
		\centering
		\includegraphics[width=\textwidth]{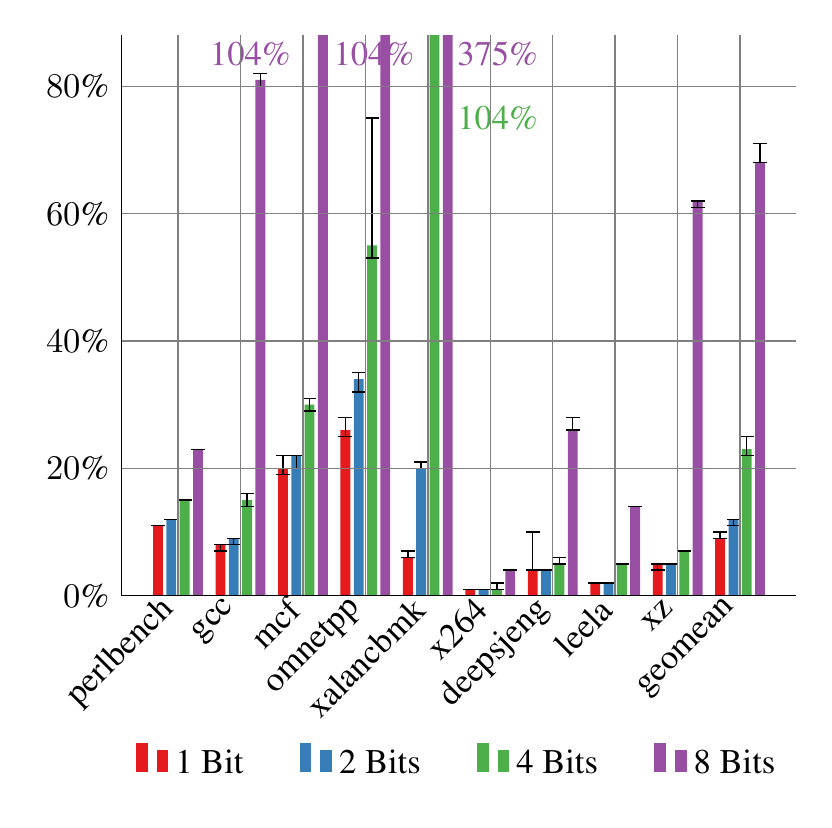}
		\caption{Tags are selected randomly.}%
		\label{fig:allocoverhead_small:rand}
	\end{subfigure}
	\begin{subfigure}{.33\linewidth}
		\centering
		\includegraphics[width=\textwidth]{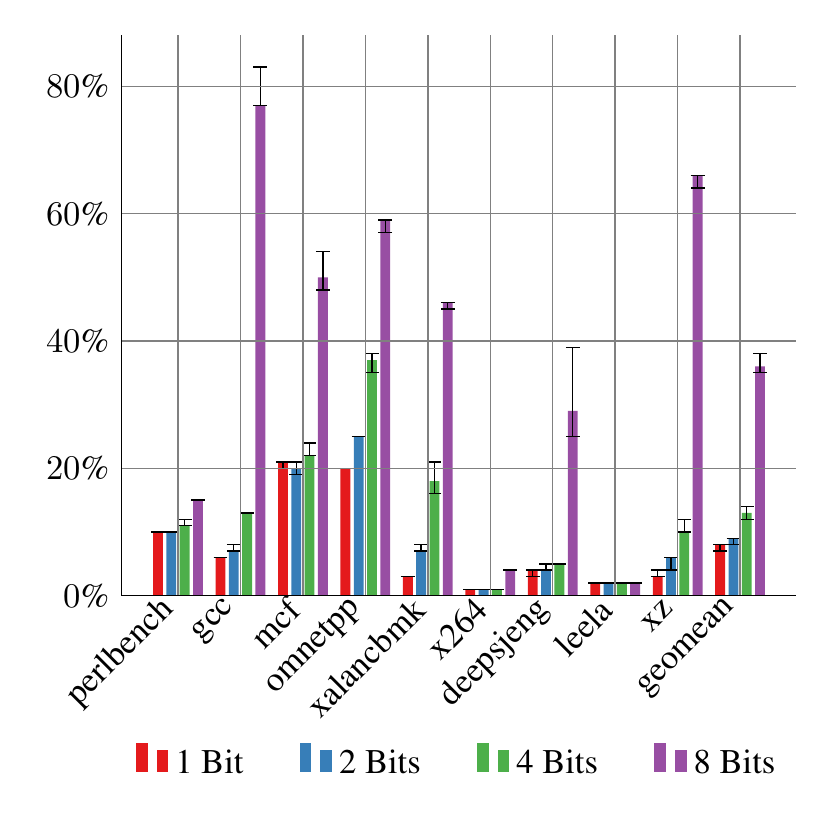}
		\caption{Tags are selected in a generational fashion.}%
		\label{fig:allocoverhead_small:gen}
	\end{subfigure}
	\begin{subfigure}{.33\linewidth}
		\centering
		\includegraphics[width=\textwidth]{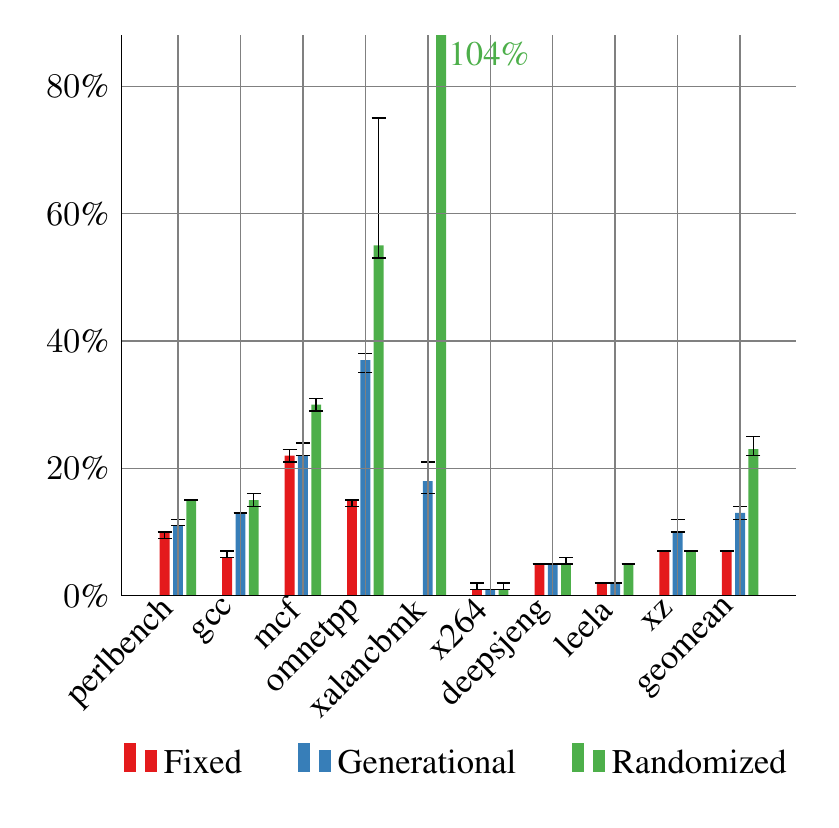}
		\caption{Comparison of tagging strategies with a tag size of 4 bit.}%
		\label{fig:allocoverhead_small:comparison}
	\end{subfigure}
	\caption{\allocname induced performance overhead with \SI{4}{\kilo\byte} pages on SPEC CPU2017. Different tag sizes and tagging strategies are compared to an unmodified \emph{mimalloc} (THP disabled) as baseline.}%
	\label{fig:allocoverhead_small}
\end{figure*}

\Cref{fig:allocoverhead_small} shows the overhead of \allocname with \SI{4}{\kilo\byte} pages for different tag sizes and selection strategies.
This evaluation is analogue to \Cref{fig:allocoverhead}, which measured the overhead for \SI{2}{\mega\byte} pages.
In general, the overhead of \allocname is higher for smaller pages sizes.
The worst individual benchmark (8 bit tags on xalancbmk with randomized tag selection) has an overhead of \SI{375}{\percent}, compared to only \SI{251}{\percent} with huge pages.
As a consequence, \toolname selects huge pages by default.

\section{Runtime Overhead Comparison}%
\label{appendix:fullrtoverhead}

\begin{figure}[h]
	\centering
	\includegraphics[width=.90\linewidth]{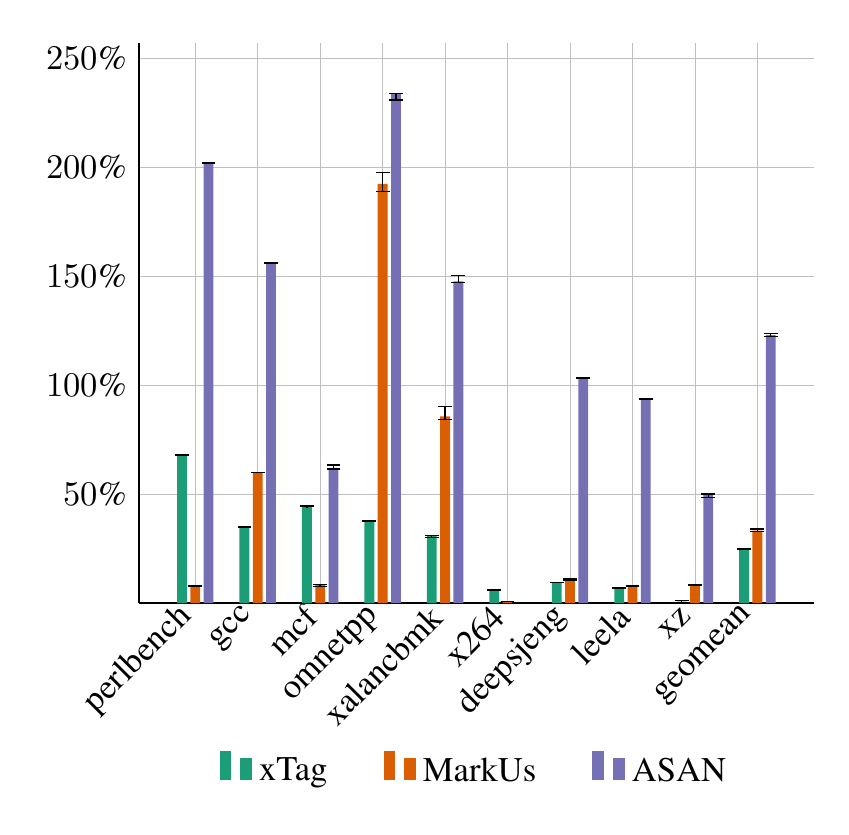}
	\caption{Runtime overhead relative to a baseline with  \emph{mimalloc} \version{v1.6.1} (THP enabled) on SPEC CPU2017 intspeed.}.
	\label{fig:appendix:RTOverheadAll}
\end{figure}

\Cref{fig:appendix:RTOverheadAll} shows the runtime overhead of all tools we measured during our benchmarking: \toolname, MarkUs, MarkUs without parallel GC, and ASAN.
We omitted ASAN from our plots in \Cref{sec:evaluation}, since ASAN is generally considered as a debugging/testing tool and not a mitigation method for production.
As such, it is not optimized to provide low overhead and, as can be seen in \Cref{fig:appendix:RTOverheadAll}, it provides the overall worst performance characteristics.
However, in contrast to \toolname\ and MarkUs, ASAN also checks for more memory safety issues, such as spatial memory safety issues and use-after-return.

\section{Serial Performance Benchmark}
\label{appendix:serialperformance}

\begin{figure}[h]
	\centering
	\includegraphics[width=.90\linewidth]{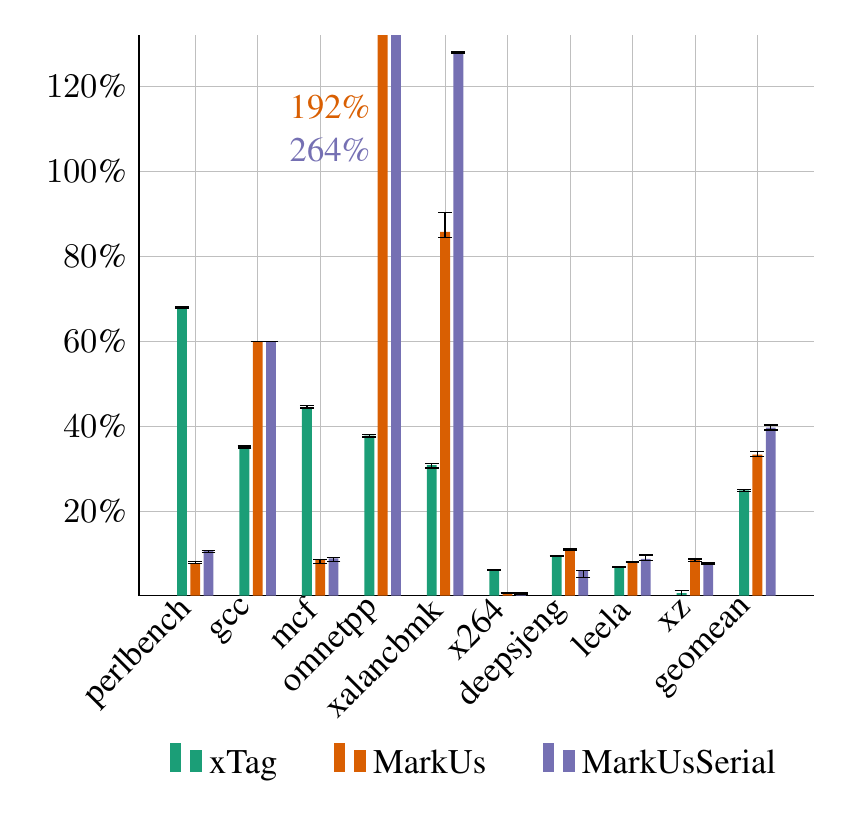}
	\caption{Runtime overhead relative to a baseline with \emph{mimalloc} \version{v1.6.1} (THP enabled) on SPEC CPU2017 intspeed.}.
	\label{fig:RTSerialOverhead}
\end{figure}

MarkUs is based on the Boehm garbage collector, a popular mark-and-sweep garbage collector.
As such, it inherits its parallel marking algorithm.
As a consequence, MarkUs can utilize idle cores during the single-core workload executed by most of the SPEC2017 intspeed benchmarks.
To offer a fair comparison with other inherently single-core mitigations such as ours, we benchmarked MarkUs both with and without the parallel garbage collection.
Our measurements for the single-core setting are designated as \emph{MarkUsSerial}, where we disabled the parallel garbage collection by configuring \texttt{--disable-parallel-mark}.
\Cref{fig:RTSerialOverhead} shows the results of our measurements and show both MarkUs variants in comparison with \toolname.
We can see that without parallel garbage collection, MarkUs' runtime overhead increases significantly, especially in workloads where MarkUs already induces a high overhead.

\section{Incompatible Functions}%
\label{appendix:ignorelist}

While in general \toolname's instrumentation is highly compatible with legacy code, there are certain pointer operations that are incompatible with \toolname's inserted instrumentation or optimizations.
To identify these functions, we run the instrumented programs and after observing a crash, we identify the offending function in the backtrace of the crash.
For the SPEC2017 benchmark, we also had to ignore several functions, with a full list shown in \Cref{tab:ignorelist}.
For Firefox, we ignore a total of 40 functions.
These functions are mainly related to the JavaScript implementation, the optimized skia graphics library, and the LZ4 decoding library.
These components exhibit highly unusual and optimized pointer usages.
One example is the \texttt{ReleaseData} function, which receives a pointer as first parameter that might be a dangling pointer.
The second parameter to this function then indicates whether the pointer actually refers to an alive object.
As such, there is no temporal memory safety violation.
However, we also cannot apply our optimizations here. 
More concretely, we cannot hoist the \uaf\ checks out of the \texttt{ReleaseData} function into the caller.

\begin{table}[h]
	\centering
	\caption{SPEC2017 functions whose pointer arguments cannot be verified in their respective callers.}%
	\label{tab:ignorelist}
	\begin{center}
		\begin{tabular}{ c }
			\toprule
			gcc                                                     \\\midrule
			\texttt{build\_function\_call\_vec}                     \\
			\texttt{operation\_could\_trap\_helper\_p}              \\
			\texttt{vn\_nary\_op\_lookup\_pieces}                   \\
			\texttt{vn\_nary\_op\_insert\_pieces}                   \\
			\toprule
			xalancbmk                                               \\\midrule
			\texttt{DeleteFunctor<>}                                \\
			\texttt{DestroyFunctor<>}                               \\
			\texttt{DestroyTable<>}                                 \\
			\bottomrule
		\end{tabular}
	\end{center}
\end{table}

\end{document}